\documentclass[traditabstract,longauth]{aa}
\usepackage[dvips]{graphicx}
\usepackage{txfonts}
\usepackage{longtable}
\usepackage{threeparttable}
\usepackage{natbib,twoopt}

\def\ms{\hbox{\,m\,s$^{-1}$}}         
\def\m2s2{\hbox{\,m$^{2}$\,s$^{-2}$}} 
\def\kms{\hbox{\,km\,s$^{-1}$}}       
\def\vsini{\hbox{$V$\,sin\,$i$}}      
\def\Msun{\hbox{$\mathrm{M}_{\odot}$}}             
\def\Rsun{\hbox{$\mathrm{R}_{\odot}$}}
\def\Mjup{\hbox{$\mathrm{M}_{\rm Jup}$}}
\def\Rjup{\hbox{$\mathrm{R}_{\rm Jup}$}}

\def\mp{M_{\rm p}}
\def\rp{R_{\rm p}}

\usepackage{lscape}
\usepackage{pdflscape}

\usepackage{array}

\newcounter{refrvno}
\DeclareRobustCommand{\refrv}[1]{%
   \refstepcounter{refrvno}%
   \therefrvno\label{#1}}

\newcounter{refrvbibno}
\DeclareRobustCommand{\refrvbib}[1]{%
   \refstepcounter{refrvbibno}%
   \therefrvbibno\label{#1}}

\newcounter{refephno}
\DeclareRobustCommand{\refeph}[1]{%
   \refstepcounter{refephno}%
   \therefephno\label{#1}}

\newcounter{refephbibno}
\DeclareRobustCommand{\refephbib}[1]{%
   \refstepcounter{refephbibno}%
   \therefephbibno\label{#1}}

\newcounter{refsysno}
\DeclareRobustCommand{\refsys}[1]{%
   \refstepcounter{refsysno}%
   \therefsysno\label{#1}}

\newcounter{refsysbibno}
\DeclareRobustCommand{\refsysbib}[1]{%
   \refstepcounter{refsysbibno}%
   \therefsysbibno\label{#1}}

\begin{document}

\title{The GAPS Programme with HARPS-N@TNG
\thanks{Tables~1, 2, 7, 8, and 9 are available in electronic form
at the CDS via anonymous ftp to cdsarc.u-strasbg.fr (130.79.128.5)
or via http://cdsweb.u-strasbg.fr/cgi-bin/qcat?J/A+A/}
}
\subtitle{XIV. Investigating giant planet migration history via improved eccentricity and mass
determination for 231 transiting planets}
\titlerunning{Investigating giant planet migration history}
\authorrunning{Bonomo et al. 2017}


\author{
A.~S.~Bonomo\inst{1} \and S.~Desidera\inst{2} \and S.~Benatti\inst{2} \and F.~Borsa\inst{3} \and S.~Crespi\inst{4} \and M.~Damasso\inst{1} 
\and  A.~F.~Lanza\inst{5} \and A.~Sozzetti\inst{1} \and G.~Lodato\inst{4} \and F.~Marzari\inst{6} \and 
C.~Boccato\inst{2} \and R.~U.~Claudi\inst{2} \and  R.~Cosentino\inst{7} \and E.~Covino\inst{8}  \and R.~Gratton\inst{2} 
\and  A.~Maggio\inst{9} \and G.~Micela\inst{9} \and  E. Molinari\inst{7} \and I.~Pagano\inst{5} \and G.~Piotto\inst{6} 
\and E.~Poretti\inst{3}  \and R.~Smareglia\inst{10}
\and L.~Affer\inst{9}
\and K.~Biazzo\inst{5}
\and A.~Bignamini\inst{10}
\and M.~Esposito\inst{8}
\and P.~Giacobbe\inst{1}
\and G.~H\'ebrard\inst{11, 12}
\and L.~Malavolta\inst{6, 2}
\and J.~Maldonado\inst{9}
\and L.~Mancini\inst{13, 1}
\and A.~Martinez~Fiorenzano\inst{7}
\and S.~Masiero\inst{9, 14}
\and V.~Nascimbeni\inst{6, 2}
\and M.~Pedani\inst{7}
\and M.~Rainer\inst{3}
\and G.~Scandariato\inst{5}
}

\institute{
INAF -- Osservatorio Astrofisico di Torino, Via Osservatorio 20, I-10025 Pino Torinese, Italy 
\and INAF -- Osservatorio Astronomico di Padova, Vicolo dell'Osservatorio 5, I-35122 Padova, Italy 
\and INAF -- Osservatorio Astronomico di Brera, Via E. Bianchi 46, I-23807 Merate (LC), Italy 
\and Dipartimento di Fisica, Universit\`a Degli Studi di Milano, Via Celoria 16, I-20133 Milano, Italy
\and INAF -- Osservatorio Astrofisico di Catania, Via S. Sofia 78, I-95123 Catania, Italy 
\and Dipartimento di Fisica e Astronomia Galileo Galilei -- Universit\`a di Padova, Vicolo dellÕOsservatorio 3, I-35122 Padova, Italy  
\and Fundaci\'on Galileo Galilei - INAF, Rambla Jos\'e Ana Fernandez P\'erez 7, E-38712 Bre\~na Baja, TF - Spain 
\and INAF -- Osservatorio Astronomico di Capodimonte, Salita Moiariello 16, I-80131 Napoli, Italy 
\and INAF -- Osservatorio Astronomico di Palermo, Piazza del Parlamento, 1, I-90134 Palermo, Italy 
\and INAF -- Osservatorio Astronomico di Trieste, via Tiepolo 11, I-34143 Trieste, Italy 
\and Institut d'Astrophysique de Paris, UMR7095 CNRS, Universit\'e Pierre \& Marie Curie, 98bis boulevard Arago, 75014 Paris, France
\and Observatoire de Haute-Provence, Universit\'e Aix-Marseille \& CNRS, F-04870 St.~Michel l'Observatoire, France
\and Max-Planck-Institut f\"ur Astronomie, K\"onigstuhl 17, D-69117 Heidelberg, Germany
\and GAL Hassin, Centro Internazionale per le Scienze Astronomiche, I-90010 Isnello, Italy
}


\offprints{A.~S.~Bonomo\\
\email{bonomo@oato.inaf.it}}

\date{Received 12 October 2016 / Accepted 1 March 2017}


\abstract{
We carried out a Bayesian homogeneous determination of the orbital parameters of 231 transiting giant planets (TGPs)
that are alone or have distant companions; we employed differential evolution Markov chain Monte Carlo methods to analyse 
radial-velocity (RV) data from the literature and 782 new high-accuracy RVs 
obtained with the HARPS-N spectrograph for 45 systems over $\sim$3~years. 
Our work yields the largest sample of systems with a transiting giant exoplanet and 
coherently determined orbital, planetary, and stellar parameters. \\
We found that the orbital parameters of TGPs in non-compact planetary systems are clearly shaped by tides raised by their host stars. 
Indeed, the most eccentric planets have relatively large orbital separations and/or high mass ratios, 
as expected from the equilibrium tide theory. 
This feature would be the outcome of planetary migration from highly eccentric 
orbits excited by planet-planet scattering, Kozai-Lidov perturbations, or secular chaos. 
The distribution of $\alpha=a/a_{\rm R}$, where $a$ and $a_{\rm R}$ are the semi-major axis and the Roche limit, 
for well-determined circular orbits peaks at 2.5; this agrees with expectations from the high-eccentricity migration (HEM), 
although it might not be limited to this migration scenario. The few planets of our sample with circular orbits and $\alpha >5$  
values may have migrated through disc-planet interactions instead of HEM. 
By comparing circularisation times with stellar ages, 
we found that hot Jupiters with $a < 0.05$~au have modified tidal quality factors $10^{5} \lesssim Q'_{\rm p} \lesssim 10^{9}$, 
and that stellar $Q'_{\rm s} \gtrsim 10^{6}-10^{7}$ are required to explain the presence of eccentric planets at the same orbital distance. \\
As a by-product of our analysis, we detected a non-zero eccentricity $e=0.104_{-0.018}^{+0.021}$ for HAT-P-29; 
we determined that five planets that were previously regarded to be eccentric or to have hints of non-zero eccentricity, 
namely CoRoT-2b, CoRoT-23b, TrES-3b, HAT-P-23b, and WASP-54b, have circular orbits or undetermined eccentricities; 
we unveiled curvatures caused by distant companions in the RV time series of HAT-P-2, HAT-P-22, and HAT-P-29;
we significantly improved the orbital parameters of the long-period planet HAT-P-17c; and 
we revised the planetary parameters of CoRoT-1b, which turned out to be considerably more inflated than previously found. 
}


 
\maketitle
%

\section{Introduction}
\label{introduction}
Despite two decades of observational efforts to discover extrasolar planets with a variety of techniques,
many fascinating issues concerning the properties and orbital evolution of 
giant planets are still open questions. Among these are the migration of hot Jupiters, the origin of 
the frequently observed spin-orbit misalignments, and 
the architecture of planetary systems with close-in giant planets. 
These planets are thought to be formed beyond the water-ice line ($a \gtrsim 1-3$~au) 
in the protoplanetary disc, where solid material is abundant because of ice condensation, 
and then migrate towards their host stars \citep{2000Icar..143....2B, 2006ApJ...648..666R}. 
Two main scenarios are usually invoked to explain the migration of hot giant planets:
disc-driven migration and high-eccentricity migration (HEM). 

The former would yield mainly circular orbits and spin-orbit alignments because of 
damping by the disc (e.g. \citealt{1980ApJ...241..425G, 2000MNRAS.315..823P, 2012ARA&A..50..211K}). 
Nevertheless, modest eccentricities $e \lesssim 0.1$ might be excited by disc-planet interactions for 
high disc surface densities and planet masses (e.g. \citealt{2013MNRAS.428.3072D}) 
or for less massive discs and planets with initial non-zero eccentricities ($e \sim 0.04$, \citealt{2015ApJ...812...94D}). 
Actually, even significant obliquities can be observed after smooth disc migration if 
the disc was primordially misaligned for instance by a distant stellar companion
\citep{2012Natur.491..418B} or chaotic star formation \citep{2010MNRAS.401.1505B}.

According to the HEM scenario, giant planets can move very close to their stars because of tidal dissipation of 
highly eccentric orbits occurring at periastron (e.g. \citealt{1996Sci...274..954R}). 
Initial high eccentricities and spin-orbit misalignments are thought to be 
produced after disc dissipation by planet-planet scattering (e.g. \citealt{2008ApJ...686..580C}), 
Kozai-type perturbations induced by a distant 
stellar or planetary companion on a highly inclined orbit (e.g. \citealt{2007ApJ...669.1298F}, \citealt{2011Natur.473..187N}),
secular dynamics in multi-planet systems (e.g. \citealt{2011ApJ...735..109W, 2017MNRAS.464..688H}), 
or a combination of these processes (e.g. \citealt{2008ApJ...678..498N}). 
Tidal dissipation inside the planet mainly acts to circularise and shrink the planetary orbit, 
while tides inside the star act to realign the orbital plane with the stellar equator \citep{2012MNRAS.423..486L}.
This explains why circularisation and spin-orbit re-alignment may procede with different timescales 
according to the different dissipation rates inside the stars and the planets (e.g. \citealt{2014ARA&A..52..171O}).

The current distribution of eccentricities of known giant planets -- with circular orbits prevalently found 
at small distances from their host stars and significant eccentricities at wider separations where tidal interactions 
are much weaker -- seem to support the high-eccentricity scenario against migration in the disc 
(see e.g. \citealt{2010ApJ...725.1995M}; \citealt{2011MNRAS.414.1278P}, hereafter P11; 
\citealt{2012MNRAS.422.3151H}, hereafter H12). 
Moreover, planets that migrated from highly eccentric orbits through tidal dissipation and 
underwent orbit circularisation without significant mass or orbital angular momentum loss
are expected to be found at a distance from their stars greater than or equal to twice the Roche limit 
$a_{\rm R}=2.16 \cdot R_{\rm p} \cdot (M_{\rm s}/M_{\rm p})^{1/3}$ 
(e.g. \citealt{2005Icar..175..248F, 2006ApJ...638L..45F}). 
It has recently been argued that this condition might 
also explain the sub-Jovian desert, the dearth of sub-Jupiter-mass planets on short-period orbits
(\citealt{2016ApJ...820L...8M}; however, see also \citealt{2016A&A...589A..75M}).

Further support for the HEM is likely provided by the eccentricity-metallicity relation for warm Jupiters, that is 
Jovian planets with orbital periods $10 \lesssim P \lesssim 200$~d, first noticed by \citet{2013ApJ...767L..24D}. 
They pointed out that eccentric warm Jupiters are mainly found around metal-rich stars 
because in metal-rich environments more Jovian planets can form and thus more frequent gravitational interactions among them 
may occur via scattering or secular perturbations and raise their eccentricities. 
The finding that planets with an outer companion have higher average eccentricites
than their single counterparts has been recently claimed by \citet{2016ApJ...821...89B} 
from results of a Doppler survey of non-transiting systems 
carried out with the HIgh Resolution Echelle Spectrometer (HIRES) on the Keck telescope. 

However, migration of close-in planets may also occur via interactions with the protoplanetary disc. 
Evidence for disc migration is mainly provided by 
i) giant planets orbiting very young stars and
ii) hot and warm giant planets in compact systems, 
that is tightly packed multi-planet systems with minimum mutual inclinations:

\renewcommand{\theenumi}{\roman{enumi}}

\begin{enumerate}
\item Young giant planets such as V830\,Tau\,b \citep{2016Natur.534..662D} 
with an age of $\sim$2~Myr likely migrated through interactions with 
the protoplanetary disc because HEM would usually require longer times 
(e.g. \citealt{2008ApJ...678..498N, 2015ApJ...799...27P, 2016ApJ...829..132P}). 
\item Two examples of hot giant planets in compact systems are the planetary systems Kepler-101 \citep{2014A&A...572A...2B} and 
WASP-47 \citep{2015ApJ...812L..18B}. The former is composed of an inner super-Neptune 
(Kepler-101b) with an orbital period $P=3.49$~d and an outer Earth-size planet with $P=6.03$~d. 
The latter contains four planets, with its hot Jupiter WASP-47b between an inner 
super-Earth ($P=0.79$~d) and an outer Neptune-size planet ($P=9.03$~d), and an additional long-period companion 
\citep{2016A&A...586A..93N}. The giant planets Kepler-101b and WASP-47b almost certainly underwent disc migration
otherwise the HEM would have destabilised the orbits of their close planetary companions.
However, systems like Kepler-101 and WASP-47 seem to represent the exception rather than the rule because 
high-precision space-based data from Kepler and CoRoT have revealed that the vast majority of hot giant planets 
do not have close companions (e.g. \citealt{2011ApJ...732L..24L, 2012PNAS..109.7982S}). 
The same considerations apply to warm Jupiters whose multiplicity rate appears to be significantly 
higher than hot giant planets \citep{2016ApJ...825...98H}. 
\end{enumerate}

Alternative theories to both high-eccentricity and disc migration to explain the presence of hot and warm giant planets 
in compact systems foresee in situ formation \citep{2016ApJ...829..114B, 2016ApJ...825...98H}, 
although these theories need observational corroboration \citep{2016ApJ...829..114B}.

To yield more observational constraints to theoretical models of giant planet migration
and star-planet tidal interactions, it is important to determine accurate and precise orbital and physical parameters, 
hopefully in a uniform way, for a large sample of giant planets. Among the known giant planets,
those seen in transit are the most interesting because transits allow us to determine the planet radii 
and, when combined with RV measurements, their true masses, 
which are fundamental physical quantities for investigating tidal effects (e.g. \citealt{1981A&A....99..126H}). 
This work aims precisely to perform a homogeneous analysis of orbital and physical parameters 
of the known transiting giant planets (hereafter, TGPs) published before 1 January 2016 
that do not belong to compact planetary systems.
A particular attention is given to the orbital eccentricity 
which is a key parameter for understanding planetary evolution (e.g. \citealt{2015A&A...574A..39D}).
This analysis is of particular importance for several reasons: 
\begin{itemize}
\item planet eccentricities are often fixed at zero in the discovery papers when found consistent with zero
or different from zero but with a low significance. Even though in some cases this assumption may
be justified (see e.g. Sect.~6 in \citealt{2012MNRAS.422.1988A}), it prevents us from 
determining the uncertainty on the eccentricity and, when this uncertainty is large, small but significant eccentricities
in principle cannot be excluded; 
\item radial-velocity (RV) data of some planetary systems discovered by independent groups 
and obtained with different spectrographs were never combined to improve the orbital solution; 
\item RV jitter terms (see e.g. \citealt{2005ApJ...631.1198G}) were not included as free parameters 
in the orbital fit of several known TGPs. This may lead to an underestimation of the eccentricity uncertainty 
and, in the worst cases, even to spurious eccentricities (H12); 
\item previous homogeneous analyses by P11 and H12 were limited to less than 70 systems while
 about 300 well-characterised TGPs are known today. 
\end{itemize}

\noindent
For these reasons, we undertook a homogeneous Bayesian analysis of RV data published in the literature 
and new RVs we collected for forty-five TGPs with the High Accuracy Radial velocity Planet Searcher 
for the Northern hemisphere (HARPS-N) spectrograph on the 3.6~m Telescopio Nazionale Galileo 
in La Palma island since mid-2012. 
We determine the orbital and physical parameters of 231 known TGPs in a consistent way, 
estimate upper and lower limits of their modified tidal quality factors, and, in general, 
attempt to understand better the orbital evolution/migration of close-in gaseous giant planets.
We also report improved constraints on some long-term trends caused by distant companions of 
known TGPs with the HARPS-N observations we obtained so far.

\section{Giant planet sample and radial-velocity datasets}

\subsection{Sample selection}
The sample of the TGPs for our homogeneous analysis was chosen as follows. First, we selected only 
transiting planets with masses $0.1 < \mp < 25~\Mjup$ and uncertainty on the mass
lower than $30\%$. We set the mass upper limit following \citet{2011A&A...532A..79S} 
and, to discard low-mass planets that may undergo a completely different evolution, 
we used a lower limit that is more conservative than the value of $0.3~\Mjup$ adopted by \citet{2015ApJ...810L..25H}.
Indeed, in addition to the fact that the mass range of TGPs between $0.1$ and $0.3~\Mjup$ is not well populated,
the transition between low-mass and giant planets may not occur sharply at $0.3~\Mjup$ 
(see Fig.~2 in \citealt{2015ApJ...810L..25H}).
We are interested in TGPs that are alone or have distant companions (if any) with 
orbital separations from the inner planet greater than 0.5~au.  
So TGPs with $\mp > 0.1~\Mjup$ in compact multi-planet systems, such as 
the already mentioned WASP-47 and Kepler-101 systems, 
but also Kepler-9 \citep{2010Sci...330...51H, 2014A&A...571A..38B} and 
Kepler-89 \citep{2013ApJ...768...14W}, were excluded.
We took into account the TGPs satisfying the aforementioned criteria 
that were published before 1 January 2016 and found 235 systems from which 
we excluded four systems, namely HAT-P-44, HAT-P-46, Kepler-14, and WTS-1. 
This yields 231 systems for our sample which are listed 
in Table~\ref{table_rvdata}. 

The HAT-P-44 and HAT-P-46 systems were excluded because, after preliminary analyses of the HIRES RV data, 
we noticed their orbital parameters strongly depend on those of their outer companions, but the latter 
are not well constrained by the HIRES RVs \citep{2014AJ....147..128H}. 
We removed Kepler-14 from our catalogue because the RV semi-amplitude cannot be simply derived 
by the orbital fit (Sect.~\ref{data_analysis}), but requires an ad hoc correction 
for the contamination by the visual stellar companion \citep{2011ApJS..197....3B}. 
Finally, we noticed that some of the epochs of the RV data of WTS-1 listed in the discovery paper 
\citep{2012MNRAS.427.1877C} are incorrect\footnote{Unfortunately, the leading author  
was unable to recover the correct epochs even though the reported orbital phases in Table~8 of 
\citet{2012MNRAS.427.1877C} should not be affected by errors.}. For this reason, we excluded it as well.

\subsection{HARPS-N radial-velocity data}
As mentioned in Sect.~\ref{introduction}, we obtained new RVs for 
45 of the selected 231 systems with the fibre-fed cross-dispersed HARPS-N echelle spectrograph on the
Telescopio Nazionale Galileo \citep{2012SPIE.8446E..1VC, 2014SPIE.9147E..8CC}. 
These data were obtained by the Global Architecture of Planetary Systems (GAPS) 
Consortium \citep{2015arXiv150903661P}.
The high-resolution ($R=115\,000$) HARPS-N spectrograph is located in a vacuum vessel which ensures temperature 
and pressure stability and thus provides very accurate RV measurements 
with instrumental drifts $< 1 \ms$ per night.

These 45 systems were chosen from the list of dwarf stars with TGPs
known at the time of the start of our programme (April 2012) 
that are brighter than $V=12.0$ and further north than $-20$~deg. 
Very fast rotators or well-studied hosting stars that already had many RV measurements 
were excluded.
A few individual targets were added to the sample during the
execution of the programme for RV monitoring after dedicated Rossiter-McLaughlin observations 
or for other specific reasons of interest.

At least seven RVs were taken for each of the 45 systems; for some of them we obtained 
more than 30 RV measurements. These observations are spread over $\sim 2.5-3$~yr
and were gathered with typical exposure times of 15~min with a simultaneous Thorium-lamp spectrum
for stars with $V < 10.5$ or in obj\_AB observing mode, that is with fibre A on target and fibre B on sky, for fainter stars.
Indeed, the simultaneous Thorium lamp does not provide a real improvement in the accuracy of RVs for $V \gtrsim 10.5$. 
Radial velocities were extracted from the high-resolution spectra with the standard online pipeline that
performs a weighted cross-correlation with the numerical mask closer to the stellar spectral type
\citep{2002A&A...388..632P}. Their typical photon-noise errors range from 2 to 10~$\ms$, 
depending on the stellar brightness and projected rotational velocity $\vsini$ \citep{2009A&A...505..853B}.
The RVs of HAT-P-2 and XO-3, which are moderately fast rotators, were obtained 
by fitting the cross-correlation function with a Gaussian function over a window of 50~$\kms$ instead of the default value
of 20~$\kms$. The reprocessing was performed using the Yabi interface at the Trieste Observatory 
\citep{2015A&A...578A..64B}.

In total, we collected 782 HARPS-N RVs, which are listed in Table~\ref{table_new_rv} 
along with their epochs given in $\rm BJD_{UTC}$ and photon-noise uncertainties. 
Some of them have already been published such as those acquired for 
Qatar-1 \citep{2013A&A...554A..28C}, HAT-P-18 \citep{2014A&A...564L..13E},
XO-2N \citep{2015A&A...575A.111D}, TrES-4 \citep{2015A&A...575L..15S}, and KELT-6 \citep{2015A&A...581L...6D}. 
The RVs of Qatar-1 and HAT-P-18 in Table~\ref{table_new_rv} are 
slightly different from those reported in \citet{2013A&A...554A..28C} and 
\citet{2014A&A...564L..13E} because they were extracted with the 
latest version of the HARPS-N pipeline (version 3.7), not available at the time
of the original publications. 
We derived the RVs of HD\,17156 from the HARPS-N spectra that were collected by \citet{2015ApJ...811L...2M} to search for 
possible variations in the $\log{R^{'}_{HK}}$ activity indicator from periastron to apoastron; 
for this reason, these data have a specific sampling, with many observations being concentrated in four nights 
close to periastron or apoastron (see Table~\ref{table_new_rv}).

\begin{table} 
\centering
\caption{HARPS-N radial-velocity measurements and associated photon-noise uncertainties.}   
\begin{tabular}{l c c c}
\hline
\hline
Target & Epoch 						& RV      & $\sigma_{RV}$ \\
	   & $\rm BJD_{\rm UTC} - 2,450,000$	& $\ms$ & $\ms$ \\
\hline
HAT-P-1 & 6147.708176  &   -2627.69 & 1.53 \\
HAT-P-1 & 6566.590673  &   -2679.27 & 1.77 \\
... & ... & ... & ... \\
\hline
\hline
\end{tabular}
\begin{tablenotes}
{\item{\raggedleft Notes. HARPS-N RV data are available at the CDS. 
A portion is shown here for guidance regarding its form and content. }}
\end{tablenotes}
\label{table_new_rv}
\end{table}

\subsection{Literature radial-velocity data}
For each system, we used the available RV datasets published in the literature before 1 January 2016
with at least four RV observations at different orbital phases and excluded all the in-transit measurements 
because the Rossiter effect was not modelled along with the orbital fit (Sect.~\ref{data_analysis}). 
We list the total number of RVs used, their timespan, and the number of independent datasets 
for each of the 231 host stars in Table~\ref{table_rvdata}.
The maximum number of independent RV datasets for a single system is five (WASP-14).

In a few cases, considerably noisy RV datasets were not used when 
they do not yield any improvement in the orbital solution and 
more precise data taken with other spectrographs are available. 
In two cases, such as Qatar-1 and TrES-4, the literature RV measurements of 
\citet{2011MNRAS.417..709A} and \citet{2007ApJ...667L.195M} were not included because they 
yield RV semi-amplitudes that are inconsistent with the HARPS-N data 
(see \citealt{2013A&A...554A..28C} and \citealt{2015A&A...575L..15S} for Qatar-1 and TrES-4, respectively).

\subsection{Additional radial-velocity data}
For the present work we used an updated set of RVs for Qatar-2, 
consisting of 42 Tillinghast Reflector Echelle Spectrograph (TRES) RVs spanning 153 days 
from the discovery paper \citep{2012ApJ...750...84B}
after correcting for a minor bug in the barycentric correction \citep{2014ApJ...782..121B},
plus 27 new TRES RVs spanning 316 additional days (C.~A.~Latham, private communication).
These RVs are listed in Table~\ref{table_Qat2HD189_rv}. 

We also included new HARPS data of CoRoT-9 that are presented in \citet{2017arXiv170306477B}  
and that allow us to derive more precise orbital parameters and, in particular, non-zero eccentricity.

\begin{table} 
\centering
\caption{TRES radial-velocity measurements of Qatar-2 and associated photon-noise uncertainties.}   
\begin{tabular}{l c c}
\hline
\hline
Epoch 					        & RV      & $\sigma_{RV}$ \\
$\rm BJD_{\rm UTC} - 2,450,000$	& $\ms$ & $\ms$ \\
\hline
5580.0116  &  431.38  &	  38.54	  \\
5581.0271  &  537.21  &	  33.54	  \\ 
... & ... & ... \\
\hline
\hline
\end{tabular}
\begin{tablenotes}
\item {\raggedleft Notes. Data are available at the CDS. 
A portion is shown here for guidance regarding its form and content. }
\end{tablenotes}
\label{table_Qat2HD189_rv}
\end{table}

\section{Bayesian data analysis}
\label{data_analysis}
The literature data and our new HARPS-N RV measurements were fitted with 
i) a Keplerian orbit model; 
ii) a Keplerian orbit and a long-term linear drift when residuals obtained with the simple Keplerian orbit 
show a significant ($ \geq 3\sigma$) slope caused by either an outer planetary/stellar companion or an activity cycle; 
iii) two non-interacting Keplerians with a possible long-term drift if the inner TGP has a known long-period companion, as in the case 
of HAT-P-13, HAT-P-17, WASP-8, etc.;  
iv) a Keplerian orbit and a curvature if data cover less than half of the period of the outer companion or the activity cycle, 
as for HAT-P-2, HAT-P-22, HAT-P-29, WASP-34, XO-2N (Sect.~\ref{trends_companions}).

The free parameters of model i) are
the transit epoch $T_{\rm c}$, the orbital period $P$, 
the RV semi-amplitude $K$, 
$e~{\cos{\omega}}$ and  
$e~{\sin{\omega}}$ (where $e$ and $\omega$ are the eccentricity and the argument of periastron), 
a jitter term $s_{\rm j}$ and a RV zero point for each dataset 
(which accounts for the RV offsets among different spectrographs).
The slope $\dot\gamma$ is the extra parameter in model ii), with the addition of 
the quadratic trend $\ddot\gamma$ for model iv). 
Model iii) obviously includes the orbital parameters of the outer planet as well.

The posterior distributions of the orbital parameters were obtained in a Bayesian framework
by means of a differential evolution Markov chain Monte Carlo (DE-MCMC) technique, 
which is the MCMC version of the differential evolution genetic algorithm
\citep{Braak2006}. This guarantees an optimal exploration of the parameter space and fast convergence 
through the automatic choice of step scales and orientations to sample
the posterior distributions \citep{2013PASP..125...83E}. 
For each system, a number of chains equal to twice the number of free parameters
were initialised close to the values of the orbital parameters reported in the literature and run 
simultaneously. The scales and directions of the jumping distribution for a given chain
are determined from two of the other chains that are randomly selected at each 
step according to the prescriptions given by \citet{Braak2006}. 
Then, by considering the Gaussian likelihood function given in Eq.~9
of \citet{2005ApJ...631.1198G}, a proposed step for each chain is accepted or rejected according 
to the Metropolis-Hastings algorithm.

Gaussian priors were imposed on $T_{\rm c}$ and $P$ from transit ephemeris and on the times of the secondary 
eclipses $T_{\rm e}$, by using Eq.~18 in \citet{2008ApJ...685..543J}, when these times are available 
from ground- and/or space-based photometry. 
The epochs of secondary eclipses provide important constraints on $e~{\cos{\omega}}$ that must be properly taken 
into account in the orbital fit. 
The adopted priors on $T_{\rm c}$, $P$, and $T_{\rm e}$ are reported in Table~\ref{table_ephemeris}. 
We used a maximum of three $T_{\rm e}$ measurements per target, if more than three 
observations of secondary eclipses were performed, which only concerns  
a few systems\footnote{This choice is motivated by some tests we made for TGPs with more than three observations of $T_{\rm e}$. 
Actually, just a single accurate measurement of $T_{\rm e}$ already provides a strong constraint on $e~{\cos{\omega}}$ 
so that Gaussian priors on more than three $T_{\rm e}$ observations do not really improve the accuracy and precision of $e$, 
while they often slow down the convergence of the DE-MCMC chains. 
Therefore, the maximum of three $T_{\rm e}$ mid-times was chosen as a trade-off between convergence speed 
and the amount of prior information contained in these measurements.}.
In a few cases, priors were directly imposed on $e~{\cos{\omega}}$ 
when their values instead of $T_{\rm e}$ measurements are reported in the literature (Table~\ref{table_ephemeris}). 
Uniform priors were considered for all the other orbital
parameters except for the jitter terms for which Jeffrey's priors were adopted (e.g. \citealt{2005ApJ...631.1198G}). 
The uniform prior on the eccentricity was ensured by weighting 
the stepping probability by the Jacobian of the transformation from 
$[e~{\cos{\omega}}, e~{\sin{\omega}}]$ to $[e, \omega]$ \citep{2006ApJ...642..505F}.
This is practically equivalent to fitting $[\sqrt{e}~{\cos{\omega}}, \sqrt{e}~{\sin{\omega}}]$ 
(e.g. \citealt{2011ApJ...726L..19A}), but we preferred $[e~{\cos{\omega}}, e~{\sin{\omega}}]$ since these quantities
appear directly in the expression of the time difference $\Delta T=T_{\rm e}-T_{\rm c}$ (e.g. \citealt{2008ApJ...685..543J}).

The DE-MCMC chains were stopped after achieving convergence and good mixing 
according to \citet{2006ApJ...642..505F}, in other words, as soon as 
the Gelman-Rubin statistic is lower than 1.01 and the number of 
independent steps is greater than 1000. 
Burn-in steps were identified following \citet{2013PASP..125...83E} 
and removed. The medians and the $15.86\%$ and $84.14\%$ quantiles of the posterior distributions  
are taken as the best values and $1\sigma$ uncertainties of the 
fitted and derived parameters. 
When the modes (or the medians) of the distributions of the eccentricity or the RV jitter were found to
be consistent with zero within $1\sigma$ ($2\sigma$), we provided their $1\sigma$ upper limits 
(and $2\sigma$ upper limits for the eccentricity) estimated as the $68.27\%$ ($95.45\%$) confidence intervals starting from zero.
The best-fit model and the RV residuals were visually checked 
for each system to be sure that everything in the analysis worked properly and that the residuals
did not show any clear deviation from the expected symmetric distribution.

Since eccentricities cannot take negative values, observational uncertainties will yield 
systematically $e > 0$ even for circular orbits. This is a well-known bias in estimating eccentricities,
as outlined by e.g. \citet{1971AJ.....76..544L}. To overcome the bias, these authors analytically 
derived the theoretical eccentricity probability function and found
that an eccentricity should be considered significant at a level of $95\%$ if $\hat{e} \geq 2.45~\sigma_{\rm e}$
rather than simply $\hat{e} \geq 2~\sigma_{\rm e}$, where $\hat{e}$ is the eccentricity expectation value and 
$\sigma_{\rm e}$ is the $1\sigma$ error.
Here we make use of the obtained DE-MCMC posterior distributions and 
Bayesian model selection between circular and eccentric models 
to establish which model is preferred when the eccentricity is found with a low significance, 
that is between 2 and 3.5$\sigma$. 
To estimate the model likelihoods and thus to compute the Bayes factors between the two models,
we sampled the posterior distributions obtained for both the circular and eccentric models 
with the \citet{Perrakis201454} method and its implementation as described in \citet{2016A&A...585A.134D}.
Bayes factors greater than 20 in favour of the eccentric 
model usually provide strong evidence for a non-zero eccentricity \citep{doi:10.1080/01621459.1995.10476572} 
and, in those cases, we considered the orbit as eccentric. 
In general, we found that Bayes factors greater than 20 
correspond to $\hat{e} > 3~\sigma_{\rm e}$, 
where $\hat{e}$ is computed as the median of the eccentricity posterior distribution.

Other indicators such as the Bayesian Information Criterion (BIC)  or the Akaike Information Criterion (AIC)
\citep{2007MNRAS.377L..74L} largely used in the literature (e.g. by P11 and H12) may suffer from 
several problems. First of all, they just provide a proxy for the Bayesian evidence 
because they take only the maximum likelihood into account and not its integral 
over the parameter space. Moreover, they rely on the assumption on Gaussianity or 
near-Gaussianity of the posterior distributions, which may not be respected in several cases;
they have an asymptotic behaviour as a function of the number of measurements 
and it is not clear whether this behaviour is reached with 10-20 measurements, as is the case for several systems of our sample
(however, the corrected AICc indicator should partly overcome this problem). 
Even more importantly, the BIC may often overestimate the odds ratio by a great extent, 
as shown by \citet{2016A&A...585A.134D}.

We classify the orbit of a TGP as decisively circular if its eccentricity is consistent with zero and 
$\sigma_{\rm e} < 0.050$ to use a quite stringent criterion, 
although inevitably any criterion such as this is somewhat arbitrary. 
Eccentricities consistent with zero but with larger uncertainties are considered `undetermined' (Sect.~\ref{results}).

\section{System parameters}
\label{system_parameters}
Our updated values of the RV semi-amplitude were combined 
with transit and stellar parameters from the literature to 
re-determine the physical planetary parameters, mass $\mp$, 
density $\rho_{\rm p}$, and gravity $\log{g_{\rm p}}$.
In Table~\ref{table_system_parameters} we list the adopted values of stellar and transit parameters 
and the corresponding references. 
In general, we gave preference to the discovery papers unless more recent analyses 
yielded significantly revised and/or much more precise parameters. 
We used stellar parameters derived from asteroseismology for the very few targets
for which asteroseismic analyses could be performed.
For the adopted literature values of system parameters we employed split-normal distributions, 
with the option of taking asymmetric error bars into account,
and combined them with the DE-MCMC posterior distributions of the orbital parameters.

System ages are crucial in order to understand the orbital evolution of planetary systems 
(Sect.~\ref{tidal_diagrams}) and to estimate upper and lower limits of the planetary modified tidal quality factors
for well-determined circular and eccentric orbits (Sect.~\ref{tidal_quality_factors}). 
While collecting the system parameters from the literature, we noticed strong inhomogeneity in computing stellar ages 
and a lack of age estimates (or age uncertainties) for some systems. 
System ages are usually derived via stellar evolutionary tracks by using constraints 
on the stellar density from the transit fitting and 
on the metallicity and the effective temperature of the host stars from spectral analysis 
(e.g. \citealt{2007ApJ...664.1190S}). 
However, for several planetary systems such as those discovered by the WASP/SWASP surveys, 
stellar ages have been mainly estimated through gyrochronology. 
Even though, in some cases, estimates from gyrochronology and stellar evolutionary tracks agree closely
(see e.g. \citealt{2015A&A...575A..85B}), gyrochronological ages may not be accurate, in general 
\citep{2015MNRAS.450.1787A} and because of possible spin-up of the host stars by their close-in planets 
\citep{2009MNRAS.396.1789P, 2015A&A...577A..90M}. 
For uniformity, in the case of age values derived from gyrochronology or lack of system ages (or their errors), 
we computed new stellar parameters, radius, mass, and age, and associated uncertainties,  
by using the Yonsei-Yale evolutionary tracks \citep{2004ApJS..155..667D} 
and determinations of transit stellar density, metallicity, and effective temperature
from the literature (values and references are given in Table~\ref{table_system_parameters}; see also 
\citealt{2014A&A...572A...2B, 2015A&A...575A..85B} for more details). 
This concerns 37 host stars (see Table~\ref{table_system_parameters}). 
For all these stars except three, i.e. CoRoT-1, WASP-63, and WASP-78, 
we found radii and masses in good agreement, within $2\sigma$, with the 
literature values even when the latter were computed with empirical relations 
\citep{2010A&ARv..18...67T, 2010A&A...516A..33E} instead of stellar models.
Our $R_{\rm s}$ and $M_{\rm s}$ of WASP-63 and WASP-78 derived with the Y2 evolutionary tracks 
differ by more than $2\sigma$ from the values of the discovery papers that were
instead estimated with empirical relations. 
The disagreement about the parameters of CoRoT-1 has a different origin: 
we used the updated value of stellar $T_{\rm eff}=6298 \pm 66$~K 
determined by \citet{2012ApJ...757..161T}. This value of $T_{\rm eff}$ is consistent with that 
found by \citet{2013A&A...558A.106M}, but 
is hotter than the $T_{\rm eff}$ previously reported by \citet{2008A&A...482L..17B}.  
We also used our $R_{\rm s}$ and $M_{\rm s}$ values for WASP-79 because our larger uncertainties encompass 
the solutions given in \citet{2012A&A...547A..61S} both with and without the main-sequence constraint. 
For the other thirty-three hosting stars we used the literature values of $R_{\rm s}$ and $M_{\rm s}$ 
(Table~\ref{table_system_parameters}).

For the systems for which we significantly improved the eccentricity determination 
(Sect.~\ref{eccentricities}), new combined analyses of RV data and literature transit photometry 
might be worthwhile as they could lead to slightly improved stellar parameters, and hence planetary parameters. 
However, this goes beyond the scope of the present work.

\section{Results}
\label{results}
In Table~\ref{table_orbital_parameters} we report the RV orbital parameters and their 1$\sigma$ uncertainties 
for the 231 TGPs of our sample. For eccentricities consistent with zero, we provide  
both the 1$\sigma$ and 2$\sigma$ upper limits. 
In the second column, we designate the orbit of each planet as circular (C), eccentric (E), or 
undetermined (U) on the basis of the criteria discussed in Sect.~\ref{data_analysis}.
In the second last column we report the long-term trends detected with significance greater than or equal to 
3$\sigma$ and/or indicate with `PLC' (planetary companion) and `CURV' the systems that show the presence of an outer planet 
or a curvature in the RV residuals, respectively. 
Trends and curvatures are usually caused by distant stellar or planetary companions, but in some cases 
may also be due to stellar activity cycles.
In the last column we list the RV jitter for each host star found for the dataset with the 
lowest median formal uncertainty. 
Indeed, stellar jitter, which is mostly caused by stellar activity 
(e.g. \citealt{2009A&A...495..959B, 2011A&A...527A..82D, 2011A&A...533A..44L}), 
can be more accurately estimated in RV data that are not dominated by photon noise 
or are dominated very little. 

Physical planetary parameters (mass, density, and surface gravity) 
were derived as explained in Sect.~\ref{system_parameters} and
are listed in Table~\ref{table_planet_parameters} along with the planet orbital periods and 
eccentricities; for this last value we report only the 1$\sigma$ upper limits in the case of circular 
or undetermined orbits. 

We first discuss significant differences and new discoveries with respect to the literature 
(Sect.~\ref{eccentricities}, \ref{trends_companions}, and \ref{updated_planetary_parameters}) 
and then report on ensemble analyses to investigate tidal evolution and tidal star-planet interactions
(Sect.~\ref{smaxes_rochelimits}, \ref{tidal_diagrams}, and \ref{tidal_quality_factors}).

\subsection{Eccentricities}
\label{eccentricities}
Several measures of the orbital eccentricities reported in Table~\ref{table_orbital_parameters} are 
the most accurate and precise ever obtained because they were derived by 
i) combining for the first time different RV datasets published in the literature;
ii) including for the first time priors on the epochs of secondary eclipses in the Keplerian fit 
(for instance, those recently observed by \citealt{2015ApJ...810..118K}); and 
iii) adding our new high-accuracy and high-precision HARPS-N data for 45 systems. 

\begin{figure}[t]  
\centering
\includegraphics[angle=90, width=9.2cm]{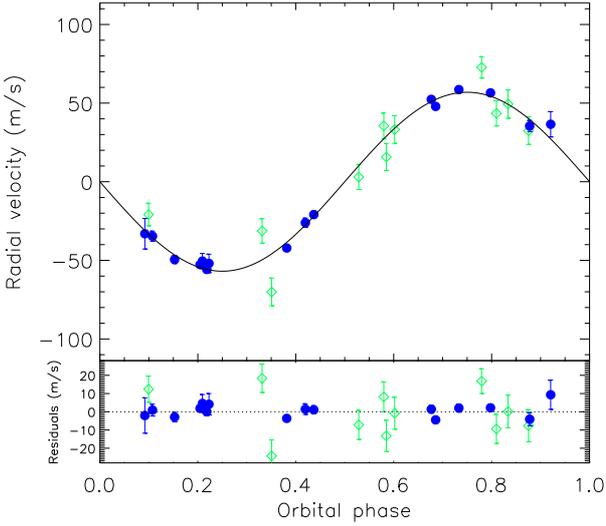}
\vspace{0.3 cm}
\caption{HARPS-N (blue circles) and SOPHIE (green diamonds) RV measurements of WASP-13 
phase-folded with the photometric ephemeris along with the best-fit Keplerian model.
We note the high-accuracy and high-precision radial velocities collected 
with HARPS-N with respect to the literature data.}
\label{figure_wasp13}
\end{figure}

An example of the improvement on the measure of the orbital eccentricity in case iii), 
especially in the absence of constraints from observations of planetary occultations, is WASP-13: 
we found $e < 0.017$, to be compared with $e=0.14 \pm 0.10$ 
by using only the literature (SOPHIE) values \citep{2009A&A...502..391S}, 
although the eccentricity was fixed to zero in the discovery paper.
Figure~\ref{figure_wasp13} shows our HARPS-N RV data of WASP-13
(blue circles) along with the SOPHIE measurements (green diamonds).


\begin{figure}[t]  
\centering
\includegraphics[angle=90, width=9.2cm]{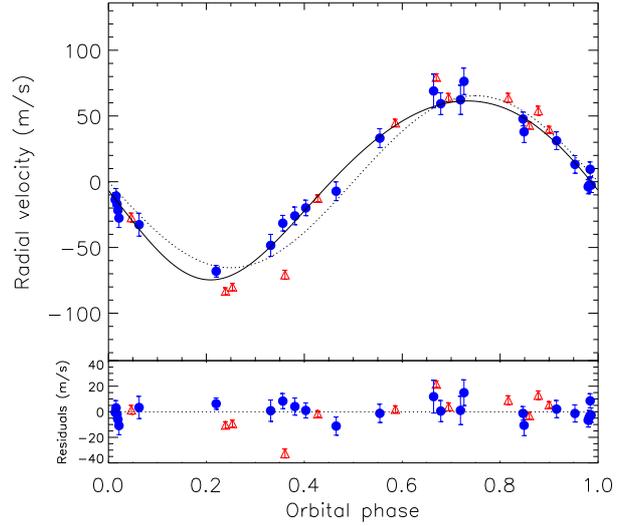}
\vspace{0.3 cm}
\caption{Phase-folded radial-velocity curve of HAT-P-29, after removing the quadratic trend (see text).
The blue circles and red triangles indicate the HARPS-N and HIRES RV measurements, respectively. 
The solid line shows the preferred eccentric orbit, 
while the dotted line displays the circular model for comparison. 
}
\label{figure_hatp29}
\end{figure}

\begin{figure}[b]  
\vspace{0.3 cm}
\hspace{-0.65 cm}
\includegraphics[angle=90, width=11.0cm]{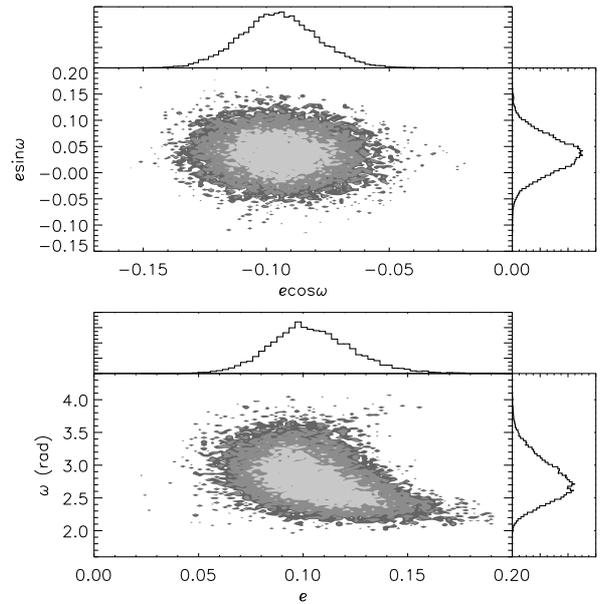}
\caption{DE-MCMC posterior distributions of $e\sin{\omega}$ vs $e\cos{\omega}$ (upper panel), and 
the derived $\omega$ vs $e$ (bottom panel), 
from our analysis of the HIRES and new HARPS-N RVs, 
showing the small but significant eccentricity of 
HAT-P-29b: $e=0.104_{-0.018}^{+0.021}$.}
\label{figure_postdistr_hatp29}
\end{figure}

With 25 high-precision and high-accuracy HARPS-N radial velocities
we found a small but significant (5.8$\sigma$) eccentricity for HAT-P-29b, that is
$e=0.104_{-0.018}^{+0.021}$ (Table~\ref{table_orbital_parameters}). 
This cannot be revealed with previous HIRES data only \citep{2011ApJ...733..116B, 2014ApJ...785..126K}.
Radial-velocity data are shown in Fig.~\ref{figure_hatp29} 
along with the eccentric and circular best fits for comparison.  
Figure~\ref{figure_postdistr_hatp29} displays the posterior distributions of $e \cos{\omega}$ versus $e \sin{\omega}$, 
and the derived $e$ and $\omega$, showing the evident eccentricity detection.
The non-zero eccentricity is not dependent on the way the long-term trend seen in HIRES RVs 
is modelled, that is with a quadratic trend (Sect.~\ref{trends_companions}) or
by fitting two independent slopes for the HIRES and HARPS-N data.


A few significant eccentricities or remarkable hints of non-zero eccentricity 
reported in the literature were found consistent with zero and/or 
were designated as `undetermined' in our analysis. 
These concern 
CoRoT-2b   ($e=0.0143_{-0.0076}^{+0.0077}$, \citealt{2010A&A...511A...3G}), 
CoRoT-23b ($e=0.16 \pm 0.02$, \citealt{2012A&A...537A..54R}), 
TrES-3b      ($e=0.170_{-0.031}^{+0.032}$, \citealt{2014ApJ...785..126K}), 
HAT-P-23b  ($e=0.106 \pm 0.044$, \citealt{2011ApJ...742..116B}), and
WASP-54b  ($e=0.067_{-0.025}^{+0.033}$, \citealt{2013A&A...551A..73F}).
As reported in Table~\ref{table_orbital_parameters}, we found a circular orbit for CoRoT-2b and TrES-3b, 
while the eccentricities of CoRoT-23b, HAT-P-23b, and WASP-54b are not well determined 
according to the adopted criteria in Sect.~\ref{data_analysis}. 
For the last object we acquired thirteen new HARPS-N RVs (Table~\ref{table_rvdata}).

In particular, the hint for a significant eccentricity of CoRoT-2b mainly came from 
the timing of the Spitzer secondary eclipse at 4.5~$\mu\rm m$ by \citet{2010A&A...511A...3G} 
that pointed to $e\cos{\omega}$ significantly different from zero. 
However, the two Spitzer observations of planetary occultation at 3.6 and 8.0~$\mu\rm m$ by \citet{2011ApJ...726...95D}
do not indicate any significant phase shift from 0.5 (see their Table 2).
The argument of periastron is unconstrained from our orbital fit including priors imposed on the secondary eclipse timings, 
thus questioning further a possible non-zero eccentricity.

The disagreement with the eccentricity of TrES-3b found by \citet{2014ApJ...785..126K} is 
remarkable, but, unlike these authors, we included all the RV measurements collected in the 
discovery paper (\citealt{2009ApJ...691.1145S}, see Table~\ref{table_rvdata}) 
where the eccentricity was fixed to zero. 
Our eccentricity $e < 0.043$ is in agreement with the timing of the secondary eclipses 
indicating $e\cos{\omega}$ consistent with zero \citep{2010ApJ...711..374F, 2010ApJ...718..920C}. 

More RV observations are required to unveil possible small but non-zero eccentricities for 
CoRoT-23b, HAT-P-23b, and WASP-54b. The same obviously applies to other systems with undetermined eccentricities, 
especially those with relatively long orbital periods ($P \gtrsim 7$~days) 
such as Kepler-39b and Kepler-74b \citep{2015A&A...575A..85B}.

\begin{figure}[h]  
\includegraphics[angle=90, width=9.2cm]{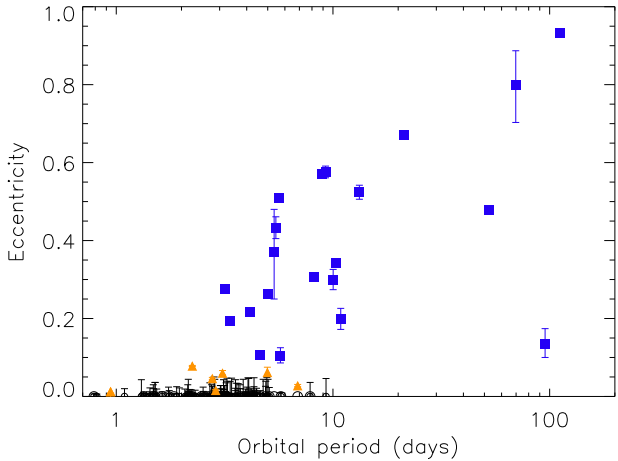}
\caption{Period-eccentricity diagram for the 123 TGPs in our sample with well-determined 
eccentricities, hence systems with undetermined eccentricities in Table~\ref{table_orbital_parameters} are not shown. 
Black empty circles refer to circular orbits, orange triangles indicate small ($e < 0.1$) but significant eccentricities,
and blue squares $e \ge 0.1$.}
\label{figure_ecc_period}
\end{figure}

Figure~\ref{figure_ecc_period} shows our derived eccentricities as a function of the 
orbital period for the 123 TGPs of our sample with well-determined eccentricities. 
Black empty circles show circular orbits, orange triangles small ($e < 0.1$) but significant eccentricities, 
and blue squares higher ($e \ge 0.1$) eccentricities.


\subsection{Long-term trends and constraints on long-period companions}
\label{trends_companions}
Seven of the 45 systems we followed up with HARPS-N were known to show 
RV drifts attributed to outer companions or stellar activity cycles:
HAT-P-2, HAT-P-4, HAT-P-7, HAT-P-22, HAT-P-29, WASP-11/HAT-P-10, and XO-2N.
Their slopes were discovered or better characterised 
with the HIRES long-term monitoring by \citet{2014ApJ...785..126K}. 

\begin{figure}[t]  
\centering
\includegraphics[angle=90, width=10cm]{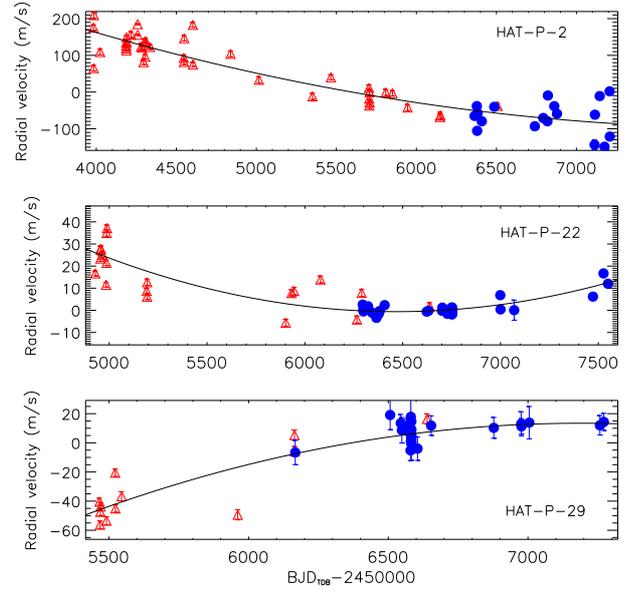}
\vspace{0.3 cm}
\caption{Radial-velocity residuals of HAT-P-2 (top), HAT-P-22 (middle), and HAT-P-29 (bottom) 
after removing the orbital signal of their inner giant planets (HAT-P-2b, HAT-P-22b, and HAT-P-29b).     
Red triangles and blue circles indicate the HIRES and HARPS-N measurements. 
The SOPHIE RVs of HAT-P-2 are not displayed here because of their relatively short timespan, 
although they were used for the orbital fit (Table~\ref{table_rvdata} and \ref{table_orbital_parameters}).
Formal error bars that are lower than symbol sizes cannot be seen.}
\label{figure_curvatures}
\end{figure}

With our HARPS-N data, we are able to confirm the trends of HAT-P-4, HAT-P-7, and WASP-11
with the same slope as found with HIRES.
For two systems, namely HAT-P-22 and XO-2N, we see different or inverted slopes and thus 
evidence for a curvature in the RV residuals, after subtracting the inner planet signal
(see Fig.~\ref{figure_curvatures}, middle panel, for HAT-P-22 and Fig.~11 in \citealt{2015A&A...575A.111D} for XO-2N). 
We do not detect any significant trend in the HARPS-N RVs of HAT-P-2 and HAT-P-29, 
which is consistent with the RV drifts found by \citet{2014ApJ...785..126K} if our data sampled 
the maximum (HAT-P-29) or minimum (HAT-P-2) of the long-term RV modulation.
The literature and HARPS-N RV data of HAT-P-2 and HAT-P-29 were thus fitted 
by including a quadratic long-term trend (see Fig.~\ref{figure_curvatures}). 

We report in Table~\ref{table_curvatures} the values of the coefficients of the curvature for HAT-P-2, HAT-P-22, and HAT-P-29 
as well as the resulting improved constraints on the orbital parameters of their outer companions,
which were derived by following the same procedure as \citet{2011AJ....142...95K} 
and using their Eqs.~1, 3, and 4.
In \citet{2015A&A...575A.111D} we already discussed the curvature of XO-2N in the HARPS-N data 
and found that it might also be due to a stellar activity cycle because of a significant correlation 
between the RV residuals and the activity index $\log(R^{'}_{HK})$ (see their Figs.~12 and 13).
We did not notice any such correlation for the other systems showing long-term trends in our data, that is
HAT-P-4, HAT-P-7, HAT-P-22, and WASP-11, which would indicate that these trends are 
due to outer companions. The same applies to HAT-P-2 and HAT-P-29 given that 
no slopes were seen in the S-index time series by \citet{2014ApJ...785..126K}.


\begin{table*}
\centering
\caption{Parameters of the curvatures shown in Fig.~\ref{figure_curvatures} and derived constraints 
on the orbital period, RV semi-amplitude, and minimum mass of the outer companions HAT-P-2c, HAT-P-22c, and HAT-P-29c.
The reference time $t_{\rm pivot}$ chosen as the average epoch of the RV time series, and the linear $\dot\gamma$ and 
quadratic $\ddot\gamma$ trends are related through Eq.~1 in \citet{2011AJ....142...95K}. }   
\begin{tabular}{l c c c c c c}
\hline
\hline
Name & $t_{\rm pivot}$ 					& $\dot\gamma$			& $\ddot\gamma$ 	  		& 	$P$		& $K$	&  $M \sin{i}$   	  \\
	  &  $\rm BJD_{\rm TDB}$ - 2,450,000	&  $ \rm m s^{-1} day^{-1}$	& $\rm m s^{-1} day^{-2}$    	&     years		& $\ms$	&  $\Mjup$ \\
\hline	
HAT-P-2c 	& 5595.65816 & $-0.077_{-0.012}^{+0.011}$ & $3.04_{-1.85}^{+1.39}\mbox{\sc{e}-05}$ & $\ge 49.2$ & $\ge 249$ & $\ge 39.5 $ \\ 
HAT-P-22c	& 5612.28617 & $-0.0052_{-0.0019}^{+0.0017}$	& $2.26\mbox{\sc{e}-05}\pm 3.0\mbox{\sc{e}-06}$ & $\ge 20.8$ & $\ge 32.9$ & $\ge 3.0$ \\
HAT-P-29c	& 5816.36464 & $0.0328 \pm 0.0064$ & $-4.0\mbox{\sc{e}-05} \pm 1.4\mbox{\sc{e}-05}$ & $ \ge 20.9$ & $\ge 59.3$ & $\ge 5.4$ \\ 
\hline
\hline
\end{tabular}
\label{table_curvatures}
\end{table*}

\begin{table*}
\centering
\caption{Improved parameters of HAT-P-17c.}   
\begin{tabular}{l c c}
\hline
Orbital Parameters & This work & \citet{2013ApJ...772...80F} \\
\hline
Orbital period $P$ [days] & $3972_{-146}^{+185} $   & $5584^{+7700}_{-2100}$ \\
Inferior conjunction epoch $T_{ \rm 0} [\rm BJD_{TDB}-2,450,000$] & $4236_{-23}^{+20}$  & $4146^{+100}_{-170}$\\
Orbital eccentricity $e$  &  $0.295 \pm 0.021$  & $0.39^{+0.23}_{-0.17}$ \\
Argument of periastron  [deg] $\omega$  &  $183.3 \pm 5.1$ & $181.5^{+5.3}_{-6.7}$  \\
Radial velocity semi-amplitude $K$ [\kms] & $42.95 \pm 0.77$ & $48.8^{+9.9}_{-6.4}$ \\
Minimum mass $M \sin{i}$ [\Mjup ]  &  $2.88 \pm 0.10$ & $3.4^{+1.1}_{-0.7}$ \\
Orbital semi-major axis $a$ [au] & 4.67 $\pm$ 0.14 & $5.6^{+3.5}_{-1.4}$ \\
\hline
\hline
\end{tabular}
\label{table_hatp17}
\end{table*}

Concerning drifts and trends that we determined only with literature RVs, 
most of which are due to outer companions,
we found values in good agreement with previous findings with the exception of HAT-P-56 and WASP-99 for which 
our analysis indicates the presence of unreported drifts with a significance of 4.0$\sigma$ and 4.6$\sigma$, respectively
(see Table~\ref{table_orbital_parameters}). 
However, more data are certainly required to establish whether these drifts are caused by outer companions or
have instrumental origin. The trend of WASP-99 is mainly driven by the first datapoint, which is $\sim 50~\ms$ 
below the other 19 measurements collected by \citet{2014MNRAS.440.1982H}. 

\begin{figure}  
\centering
\includegraphics[angle=90, width=9.2cm]{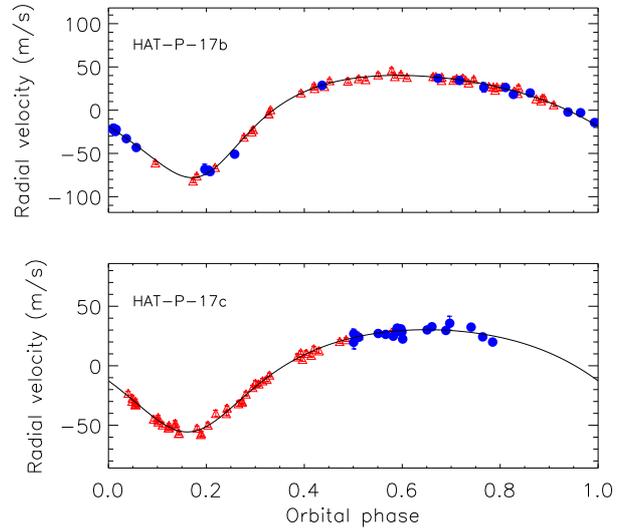}
\vspace{0.3 cm}
\caption{Keplerian best fits to the RV curves of the inner giant planet HAT-P-17b (top) and 
its outer long-period ($P\sim 4000$~d) companion HAT-P-17c (bottom).
HIRES and HARPS-N data are shown with red triangles and blue circles. 
Formal error bars of the RVs are usually smaller than symbol sizes.
}
\label{figure_hatp17}
\end{figure}

With our HARPS-N data we could detect and/or characterise better the two long-period companions
HAT-P-17c and KELT-6c of the inner giant planets HAT-P-17b and KELT-6b, respectively. 
We derived the orbital parameters of KELT-6c previously in \citet{2015A&A...581L...6D} and
we report in Table~\ref{table_hatp17}  the improved parameters of HAT-P-17c. The precision on 
its orbital period, eccentricity, and mass is at least two times better than \citet{2013ApJ...772...80F}
thanks to the extended orbital coverage. 
In Fig.~\ref{figure_hatp17} we show the Keplerian models of both HAT-P-17b and HAT-P-17c
overplotted on the HIRES (red triangles) and HARPS-N (blue circles) radial velocities.


The derived orbital parameters of the outer companions with a complete (or nearly complete) RV orbital coverage such as 
HAT-P-13c, Kepler-424c, Kepler-432c, and WASP-8c are fully consistent with 
the published values. As in \citet{2014ApJ...785..126K}, we had to fit a circular orbit for WASP-8c 
to reach convergence of the DE-MCMC chains. 
We tried to do the same for WASP-34, but failed to reach convergence and proper mixing of the DE-MCMC chains 
given the poor orbital coverage of the outer companion (see Fig.~9 in \citealt{2014ApJ...785..126K}) and thus, 
unlike these authors, we fitted a curvature.

\subsection{Updated stellar and planetary parameters}
\label{updated_planetary_parameters}
The derived masses, densities, and surface gravities, of the 231 TGPs 
are listed in Table~\ref{table_planet_parameters}. 
Our uncertainties on planetary masses are comparable with 
the literature error bars, with differences typically less than 10\% for 37\% of the considered TGPs, 
while they are lower (higher) by $>10\%$ in 35\% (28\%) of the remaining cases.
Lower error bars come from the fact that different RV datasets have been combined for the first time  
to obtain a more precise orbital solution and hence a more precise planetary mass from the improved RV semi-amplitude.
Higher uncertainties may be due to 
i) the inclusion of RV jitter terms in the orbital fit when not previously taken into account and/or
ii) the choice of letting the eccentricity vary in the orbital fit instead of adopting a circular orbit, 
as often done when the eccentricity is not well constrained but is compatible with zero. 
However, in general, the uncertainty on the eccentricity must be propagated to that of the other orbital parameters.
Therefore, both i) and ii) provide more realistic uncertainties on the orbital hence planetary parameters. 
Nonetheless, the absence of any significant eccentricity for $\mp/M_{\rm s} \lesssim 0.002$ and 
$a/\rp \lesssim 100$ (see Sect.~\ref{tidal_diagrams} and Fig.~\ref{fig_tidal_diagram}) would indicate
that assuming a circular orbit for this range of parameters is reasonable when 
RV data do not allow us to constrain well the orbital eccentricity.

As discussed in Sect.~\ref{system_parameters}, 
we used updated values of the radius and mass of four planet-hosting stars, 
namely CoRoT-1, WASP-63, WASP-78, and WASP-79 (see Table~\ref{table_system_parameters}). 
The new stellar radius and mass imply revised planetary parameters for the 
hosted planets, although only those of CoRoT-1b, and in particular its radius, 
significantly differ from the literature values. 
Indeed, CoRoT-1b is larger, hence more inflated, than originally found by \citet{2008A&A...482L..17B}. 
Its radius, mass, and density are found to be
$\rp=1.715 \pm 0.03~\Rjup$, $\mp=1.23 \pm 0.10~\Mjup$, and $\rho_{\rm p}=0.302_{-0.028}^{+0.031}~\rm g\;cm^{-3}$,
which can be compared with the previous parameters: 
$\rp=1.49 \pm 0.08~\Rjup$, $\mp=1.03 \pm 0.12~\Mjup$, and $\rho_{\rm p}=0.38 \pm 0.05~\rm g\;cm^{-3}$
\citep{2008A&A...482L..17B}. 
The reason for the larger $\rp$ lies in the fact that we used a hotter stellar $T_{\rm eff}$ 
than \citet{2008A&A...482L..17B} (see Sect.~\ref{system_parameters}); this implies a slightly higher stellar mass and 
a significantly larger stellar radius hence planetary radius from the transit depth.

\subsection{Semi-major axes and Roche limits}
\label{smaxes_rochelimits}
We find that only $4\%$ of the TGPs in our sample do not fulfil the condition $a \ge 2a_{\rm R}$, 
as predicted by HEM in the absence of mass and angular momentum loss.
Indeed, only 9 planets out of 231 have 
semi-major axes that are significantly lower than $2a_{\rm R}$, 
after taking the uncertainties on both $a$ and $a_{\rm R}$ into account: 
CoRoT-1b, HAT-P-32b, Kepler-41b, TrES-4b, WASP-12b, WASP-19b, WASP-52b, WASP-103b, and WTS-2b. 
However, four of them, Kepler-41b, WASP-52b, WASP-103b, and WTS-2b, 
have undetermined eccentricities, although their orbits are likely circular.
The planets in our sample with the lowest ratio $\alpha=a/a_{\rm R}$ are
WASP-12b ($\alpha=1.21$), WASP-19b ($\alpha=1.27$), and WASP-103b ($\alpha=1.31$).

\begin{figure}[b]  
\centering
\includegraphics[width=6.5cm, angle=90]{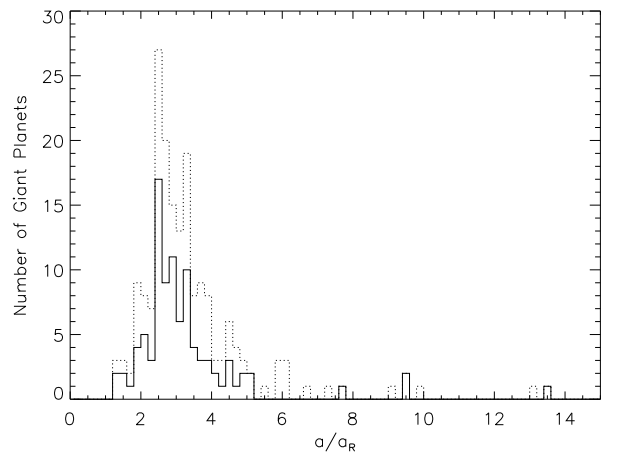}
\caption{Distributions of  $\alpha=a/a_{\rm R}$, where $a_{\rm R}$ is the Roche limit, 
for well-determined circular orbits (solid line) 
and both circular orbits and undetermined eccentricities 
(dashed line). The impact of transit probability on the distribution for $\alpha \le 5$ 
appears to be practically negligible (see text for more details).}
\label{alpha_histogram}
\end{figure}

Figure~\ref{alpha_histogram} shows the distributions of $\alpha=a/a_{\rm R}$ 
for well-determined circular orbits (solid line) and planets with both circular orbits 
and undetermined eccentricities (dotted line). Both distributions peak at 2.5. 
A relevant issue concerns whether and how the transit probability ($R_{\rm s}/a$ for circular orbits) 
affects the $\alpha$ distribution because the closer the planet, 
the higher the probability seen in transit (hence discovered). 
To evaluate the impact of the transit probability on the $\alpha$ distribution, 
we considered the distribution for circular orbits (solid line) and $\alpha \le 5$ 
that encompasses the vast majority of the circular planets of our sample. 
For each bin of 0.20 in $\alpha$, we weighted the frequency of giant planets in the bin by the 
transit probability, by adding up the inverses of the transit probabilities ($R_{\rm s}/a$) of the 
planets falling in that bin. 
The obtained normalised distribution is statistically indistinguishable from that shown 
in Fig.~\ref{alpha_histogram} (solid line) and therefore is not displayed. 
This can be understood by noticing that planets with similar semi-major axes 
(hence transit probabilities for similar $R_{\rm s}$) may fall in different $\alpha$ bins 
because of different $a_{\rm R}$ values depending on both $\rp$ and $(M_{\rm s}/\mp)^{1/3}$.

Planets with circular orbits and relatively large $\alpha$ values ($\alpha > 5$), 
such as CoRoT-3b ($\alpha=13.9$), CoRoT-27b ($\alpha=9.4$), 
WASP-99b ($\alpha=7.7$), and WASP-106b ($\alpha=9.4$), 
appear as possible outliers with respect to the inner distribution (see Fig.~\ref{alpha_histogram}). 
These planets might have undergone a different migration history, for instance 
smooth disc migration instead of HEM (see Sect.~\ref{tidal_diagrams}).

\subsection{Tidal diagrams}
\label{tidal_diagrams}
\begin{figure}[t]  
\centering
\includegraphics[width=6.5cm, angle=90]{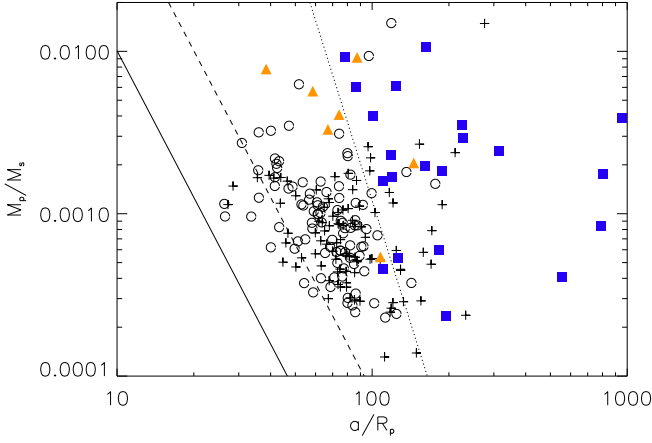}
\caption{Updated tidal diagram of the 231 transiting giant planets considered in this study. 
Empty circles show the position of giant planets with well-determined circular orbits, that is with eccentricities consistent with zero and 
$1\sigma$ uncertainty $\sigma_{e}<0.05$ (planets designated with `C' in the second column of Table~\ref{table_planet_parameters}); 
crosses indicate planets with undetermined eccentricities (indicated with `U' in Table~\ref{table_planet_parameters}) 
that are usually consistent with zero (circular orbits) but have large uncertainties $\sigma_{e}>0.05$; 
orange triangles display significant but small eccentricities, i.e. $e < 0.1$; and blue squares $e \ge 0.1$. 
Solid and dashed lines show the position of a planet with $R_{\rm p}=1.2~R_{\rm Jup}$ 
and semi-major axis $a=a_{\rm R}$ and $a=2a_{\rm R}$, $a_{\rm R}$ being the Roche limit. 
The dotted line displays the 1\,Gyr circularisation isochrone for $P=3$~d, 
$Q'_{\rm p}=10^6$, and $e=0$. 
 }
\label{fig_tidal_diagram}
\end{figure}

`Tidal diagrams' are very useful in order to understand the impact of tidal star-planet interactions 
on the planetary orbital parameters (see e.g. P11). 
They show the mass ratios $M_{\rm p}/M_{\rm s}$ as a function of $a/R_{\rm p}$ because 
the circularisation time $\tau_{e}$ scales as $\tau_{e} \propto (M_{\rm p}/M_{\rm s}) \cdot (a/R_{\rm p})^5$ 
(e.g. \citealt{1966Icar....5..375G}). 
The tidal diagram containing the 231 giant planets of our sample is displayed in 
Fig.~\ref{fig_tidal_diagram} where 
well-determined circular orbits, undetermined eccentricities, 
small but significant eccentricities  $e < 0.1$, and eccentricities $e \ge 0.1$
are indicated with empty circles, crosses, orange triangles, and blue squares, respectively.
The solid and dashed lines show the position of a planet with $R_{\rm p}=1.2~R_{\rm Jup}$, 
which represents the median of the planetary radii of our sample,
and semi-major axis $a=a_{\rm R}$ and $a=2a_{\rm R}$, respectively. For comparison with P11, 
the dotted line displays the 1\,Gyr isochrone for orbital circularisation 
for $P=3$~d (corresponding to the pileup of hot Jupiters; e.g. \citealt{2009ApJ...693.1084W}), 
$Q'_{\rm p}=10^6$, and $e=0$, where $Q'_{\rm p}=3Q_{\rm p}/2k_{2}$ is the planetary modified tidal quality factor,  
$Q_{\rm p}$ is the planet tidal quality factor, and $k_{2}$ is the Love number (e.g. \citealt{1999ssd..book.....M}).
$Q'_{\rm p}$ is a parameterisation of the response of the planet's interior to tidal perturbation: 
the higher $Q'_{\rm p}$, the lower the dissipation rate of the kinetic energy of the tides inside the planet.
A similar definition applies to the stellar modified quality factor $Q'_{\rm s}$.
The 1\,Gyr isochrone for orbital circularisation displayed in Fig.~\ref{fig_tidal_diagram}  
was computed with Eq.~6 in \citet{2008ApJ...686L..29M} (hereafter M08) 
which also applies to the case of high eccentricities and assumes that tides in the planet
are mainly responsible for circularising its orbit (e.g. \citealt{2010ApJ...725.1995M}).

As previously noticed by P11 and H12, but with a much smaller sample than ours (by a factor of three),
this diagram clearly shows that the orbital parameters of TGPs in non-compact planetary systems 
are shaped by tidal interactions with their host stars. 
Indeed, all the TGPs with $e>0.1$ (blue squares) have large separations $a/R_{\rm p} \gtrsim 100$. 
The two planets with $a/R_{\rm p}<100$, namely WASP-89b and XO-3b, have high mass ratios $M_{\rm p}/M_{\rm s} > 6 \cdot 10^{-3}$, 
as expected from tidal theory. Moreover, virtually all the TGPs with $e>0.1$ 
are found on the right side of the dotted line, that is their circularisation time is longer than 1\,Gyr.
We note that the slightly eccentric planets ($e<0.1$, orange triangles) with $a/R_{\rm p}<100$ are located at the 
upper edge of the circular ones. 

We confirm the dearth of close-in circular planets with $M_{\rm p}/M_{\rm s} \gtrsim 4 \cdot 10^{-3}$. 
These massive planets likely raise tides 
in the star strong enough for angular momentum exchange and tidal decay so that they end up being engulfed by their host star 
(e.g. \citealt{2002ApJ...568L.117P}). 
However, this scarcity might also be partially explained by the higher inertia of massive planets to gravitational scattering 
towards inner orbits and/or a less efficient formation in discs that are not 
massive enough. 

It is convenient to also consider a `modified' tidal diagram where the circularisation isochrones 
do not depend on the orbital period $P$, by plotting $P \cdot M_{\rm p}/M_{\rm s}$ versus $a/R_{\rm p}$.
From the aforementioned Eq.~6 in M08 (see also \citealt{1966Icar....5..375G}) we find

\begin{equation} 
\label{eq_modified_tidal_diagram}
P  \frac{M_{\rm p}}{M_{\rm s}} = \tau_{e}  \frac{63}{2\pi} \frac{1}{Q'_{\rm p}} \left(\frac{a}{\rp}\right)^{-5}.
\end{equation}

\begin{figure}[t]  
\centering
\includegraphics[width=6.5cm, angle=90]{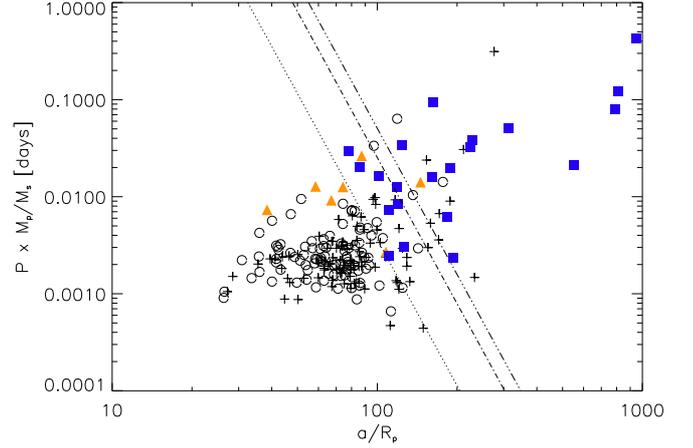}
\caption{`Modified' tidal diagram of the 231 transiting giant planets considered in this study 
with the same symbols as in Fig.~\ref{fig_tidal_diagram}. 
The dotted, dash-dotted, and dash-three-dotted lines display the 1, 7, and 14\,Gyr circularisation timescales 
for $Q'_{\rm p}=10^6$ and $e=0$, respectively. 
 }
\label{fig_modified_tidal_diagram}
\end{figure}

\noindent
While the orbital period was fixed at $P=3$~d for the tidal diagram in Fig.~\ref{fig_tidal_diagram}, 
we show for comparison $P \cdot M_{\rm p}/M_{\rm s}$ as a function of $a/R_{\rm p}$ in 
Fig.~\ref{fig_modified_tidal_diagram} along with the 1, 7, and 14\,Gyr circularisation timescales 
for $Q'_{\rm p}=10^6$ and $e=0$. 
This `modified' tidal diagram reveals even more evidently the same trends discussed above for the 
`classical' tidal diagram thanks to the independence of the isochrones on the adopted orbital period. 
Like in Fig.~\ref{fig_tidal_diagram}, all the eccentric planets in Fig.~\ref{fig_modified_tidal_diagram} with $e \ge 0.1$ (blue squares) 
are found beyond the 1\,Gyr circularisation isochrone. The four TGPs with well-determined circular orbits, $a/\rp \gtrsim 100$, and 
$P \cdot (M_{\rm p}/M_{\rm s}) \gtrsim 0.01$~d are CoRoT-3b, CoRoT-27b, WASP-99b, and WASP-106b, 
all having $\alpha > 5$ (Sect.~\ref{smaxes_rochelimits} and Fig.~\ref{alpha_histogram}). 
Their circular orbits might be primordial since they lie in a region of the `modified' tidal diagram 
where tidal circularisation is not expected to occur within 7~Gyr (or even 14~Gyr). We note that this part 
of the diagram is otherwise populated by eccentric planets (cf. Fig.~\ref{fig_modified_tidal_diagram}). 
Therefore planets with $\alpha > 5$ may have undergone disc migration instead of HEM.

\subsection{Constraints on planetary and stellar modified tidal quality factors}
\label{tidal_quality_factors}
After identifying circular and eccentric orbits, we can estimate the 
upper and lower limits of $Q'_{\rm p}$ 
by considering a tidal constant $Q'$ model (e.g. \citealt{2013LNP...861.....S, 2014ARA&A..52..171O}). 
Indeed, following M08, 
an upper limit on $Q'_{\rm p}$ can be derived from their Eq.~7, 
that is from the constraint that the circularisation time must be shorter than the system (or stellar) age 
for circular planets ($\sigma_{e}<0.05$), on the assumption that their orbits were initially eccentric.
On the contrary, lower limits on both $Q'_{\rm p}$ and $Q'_{\rm s}$ can be derived with 
Eqs.~8 and 9 in M08, by imposing that the circularisation time must be longer than the system age for eccentric orbits, 
if we assume that the eccentricity is not presently excited by a third body in the system.

These equations require the knowledge of both the stellar and planetary rotation periods. 
Stellar rotation periods as derived from ground-based or space-based photometry, 
which are reported in the seventh column of Table~\ref{table_system_parameters} when available, 
are used for approximately thirty systems. For the other systems they were estimated 
from the $\vsini$ and $R_{\rm s}$ assuming a stellar inclination of $90$~deg, 
which thus provides upper limits for the true $P_{\rm rot}$ when the stellar equator is not seen edge-on. 
This may lead to slightly overestimated lower limits of $Q'_{\rm s}$, 
but this effect is negligible in the vast majority of cases.
Concerning the planetary spin, we assume 
i) perfect synchronisation for the circular planets, which is reasonable given that 
synchronisation times are much lower than the circularisation times (e.g. \citealt{1996ApJ...470.1187R}), and 
ii) pseudo-synchronisation for the eccentric systems \citep{1981A&A....99..126H}. 

\begin{figure}[h!]  
 \centering
 \includegraphics[width=6.5cm, angle=90]{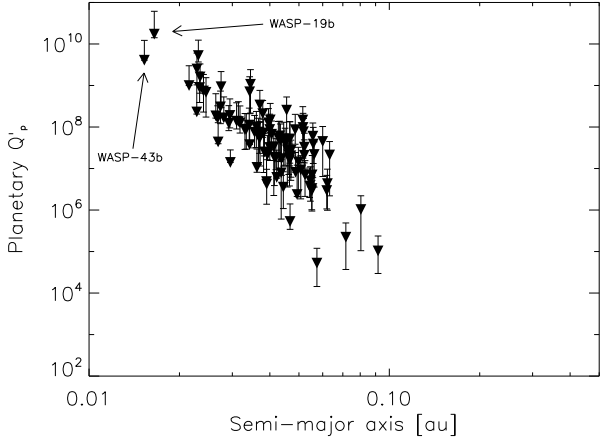}
 \caption{Upper limits of the planetary modified tidal quality factor $Q'_{\rm p}$ for clearly circular orbits.}
 \label{Qpplanet_upplimit_smaxis}
 \end{figure}

\begin{figure}[h!]  
 \centering
 \includegraphics[width=6.5cm, angle=90]{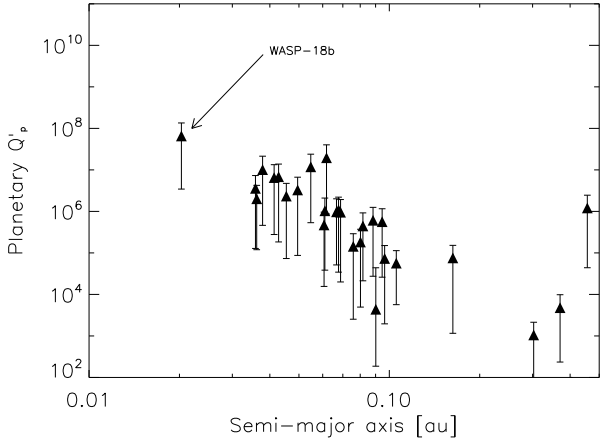}
 \caption{Lower limits of the planetary modified tidal quality factor $Q'_{\rm p}$ for eccentric orbits.}
 \label{Qpplanet_lowlimit_smaxis}
 \end{figure}

\begin{figure}
 \centering
 \includegraphics[angle=90, width=9cm]{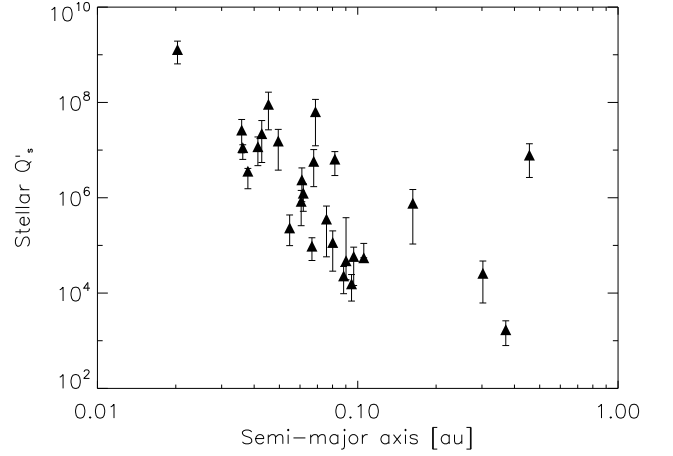}
 \caption{Lower limits of the modified tidal quality factor $Q'_{\rm s}$ of stars hosting eccentric planets.}
 \label{Qpstar_lowlimit_smaxis}
 \end{figure}

Figures~\ref{Qpplanet_upplimit_smaxis}, \ref{Qpplanet_lowlimit_smaxis}, and \ref{Qpstar_lowlimit_smaxis} 
show the upper and lower limits of $Q'_{\rm p}$, and the lower limits of $Q'_{\rm s}$, as a function of the semi-major axis. 
The large error bars are due to the propagation of the uncertainties on system ages; 
unconstrained values of $Q'_{\rm p}$ and $Q'_{\rm s}$ due to unconstrained ages in 
Table~\ref{table_system_parameters} (e.g. those spanning 0-14~Gyr at 1$\sigma$) are not shown. 
The lower limits of $Q'_{\rm p}$ also depend on the $Q'_{\rm s}$ values (see Eq.~8 in M08) and 
we adopted values of $Q'_{\rm s}$ that are ten times their lower limits: $Q'_{\rm s}= 10 \cdot Q'_{\rm s, min}$. 
This explains why the variations seen in Figs.~\ref{Qpplanet_lowlimit_smaxis} 
and \ref{Qpstar_lowlimit_smaxis} are evidently correlated.

Hot Jupiters with orbital distances typically lower than 0.05~au have 
$10^{5} \lesssim Q'_{\rm p} \lesssim 10^{9}$, but 
estimates of the $Q'_{\rm p}$ lower and upper limits are inevitably affected 
by the large uncertainties on stellar ages. 
The main trends in Figs.~\ref{Qpplanet_upplimit_smaxis} and \ref{Qpplanet_lowlimit_smaxis}
seem to indicate that the closer the planet, the higher its $Q'_{\rm p}$ 
hence its lower internal dissipation. This is also related to the differences in tidal frequencies 
as a function of the orbital distance.
Little can be said for planets at semi-major axis 
$a\gtrsim0.1$~au, given the small number of systems.

Modified tidal quality factors of stars hosting hot Jupiters with $a<0.05$~au are generally 
$Q'_{\rm s} \gtrsim 10^{6}-10^{7}$ (Fig.~\ref{Qpstar_lowlimit_smaxis}).  
The highest value $Q'_{\rm s} > 10^{8}$ in Fig.~\ref{Qpstar_lowlimit_smaxis} is that of the star WASP-18 
whose hot Jupiter at $a=0.02$~au ($P=0.94$~d) has a small but significant eccentricity: $e=0.0076 \pm 0.0010$. 
This is consistent with estimates relying on the timescales of the orbital decay of 
very hot Jupiters with $P \lesssim 1$~d (e.g. \citealt{2014ARA&A..52..171O}).

\section{Summary, discussion, and conclusions}
We carried out a homogeneous determination of the orbital parameters of 231 TGPs by analysing 
with our Bayesian DE-MCMC tool both the literature RVs and the new high-accuracy and high-precision HARPS-N data 
we acquired for 45 TGPs orbiting relatively bright stars over $\sim$3~years. 
We thus produced the largest uniform catalogue of giant planet orbital and physical parameters. 
For several systems we combined for the first time RV datasets collected with different spectrographs 
by different groups thus improving the orbital solution. 
In general, we fitted a separate jitter term for each dataset by allowing for 
different values of extra noise caused by instrumental effects and/or  
changing levels of stellar activity in different observing seasons.

This way, we uniformly derived the orbital eccentricities of TGPs that represent 
a fundamental imprint of their migration history.
We detected for the first time a significant eccentricity for HAT-P-29b. 
On the contrary, our results do not confirm any significant eccentricity for
five planets that were previously regarded to be eccentric or to have hints of 
non-zero eccentricity, namely CoRoT-2b, CoRoT-23b, TrES-3b, HAT-P-23b, and WASP-54b. 
In particular, our results favour a circular orbit for CoRoT-2b and TrES-3b.

Limiting ourselves to the 45 systems we monitored with HARPS-N, 
we confirm the RV long-term linear trends for three of them, HAT-P-4, HAT-P-7, and WASP-11, 
with the same slope as previously found \citep{2014ApJ...785..126K}. 
We report the first evidence of a curvature in the RV residuals of 
HAT-P-2, HAT-P-22, and HAT-P-29, and derived constraints on their long-period companions
from RV data only.
Moreover, our HARPS-N measurements allowed us to refine significantly the orbital solution 
of HAT-P-17c, the outer planetary companion of HAT-P-17b, thanks to the extended orbital coverage.

Our tidal diagrams clearly show how the orbital parameters of TGPs in
non-compact planetary systems are shaped by tides raised by their host stars. 
Indeed, the most eccentric planets have either relatively large orbital separations and/or 
high mass ratios, as expected from the equilibrium tide theory. 
This feature would be the outcome of planetary migration from highly eccentric 
orbits that were originally produced by planet-planet scattering, Kozai-Lidov perturbations, or secular chaos. 

The $\alpha=a/a_{\rm R}$ distribution showing that only $\sim 4\%$ of circular TGPs 
have $a < 2a_{\rm R}$ agrees with the theoretical prediction by the HEM 
that the final distances of circularised orbits must be $a \ge 2a_{\rm R}$ (e.g. \citealt{2005Icar..175..248F}).
The handful of TGPs with $a < 2a_{\rm R}$ may also be explained in the framework of
the HEM by considering the effect of tidal decay driven by tidal dissipation inside the star, 
which may account for even the lowest observed $\alpha \sim 1.2$ value \citep{2014ApJ...787L...9V}. 
The condition $a > 2a_{\rm R}$, however, may not be a peculiar imprint of the HEM and, for instance, could also be reproduced 
by migration in the disc that was stopped because of truncation of the inner disc by magnetic fields 
(e.g. \citealt{1996Natur.380..606L}).
The derived $\alpha$ distribution peaking at 2.5 represents an important observational constraint that 
theoretical models of planet migration must be able to reproduce.
The planets of our sample with circular orbits and relatively large $\alpha$ ($> 5$) values, namely
CoRoT-3b, CoRoT-27b, WASP-99b, and WASP-106b, seem to deviate from the 
inner $\alpha$ distribution. They have circular orbits although their circularisation timescale (in case of a non-zero eccentricity)
is longer than $\sim7-14$~Gyr. This would indicate that their circular orbits are likely primordial, that is they 
presumably migrated through disc-planet interactions, which tend to damp any small eccentricity, 
instead of HEM.

As previously discussed, strong evidence for disc migration is mainly provided by the discovery of giant planets around very young stars
and in compact systems. 
In addition, \citet{2015ApJ...798...66D} found a paucity of super-eccentric proto-hot Jupiters\footnote{
Super-eccentric proto-hot Jupiters are defined as highly eccentric ($e > 0.9$) giant planets 
that could become hot Jupiters through the mechanism of HEM.}
among the \emph{Kepler} sample that seems to be inconsistent with the theoretical predictions of HEM by  
\citet{2012ApJ...750..106S}. 
However, alternative explanations of this dearth are also possible in the context of HEM, for instance if 
gravitational scattering of giant planets occurs during migration of type II inside the water-ice line
(see, e.g. \citealt{2010A&A...514L...4M, 2011ApJ...732...74G}, and Sect.~4 in \citealt{2015ApJ...798...66D}). 
This may also reconcile the HEM with the apparently inconsistent occurrence of planetary companions of hot Jupiters 
inside the water-ice line \citep{2016ApJ...825...62S}.
In any case, it is difficult to explain the general properties of the tidal diagrams  
shown in Figs.~\ref{fig_tidal_diagram} and \ref{fig_modified_tidal_diagram} and discussed in Sect.~\ref{tidal_diagrams}, 
in terms of migration in the disc alone given that disc-planet interactions may only excite 
modest eccentricities $e < 0.1$ \citep{2015ApJ...812...94D}.

We estimated upper and lower limits of the planetary modified tidal quality factors $Q'_{\rm p}$ and found that 
high values (up to $10^{7}-10^{9}$) are required to explain 
the presence of the closest giant planets with $a < 0.05$~au. 
These high values of $Q'_{\rm p}$ are in agreement with the predicted very low internal dissipation in  
massive planets with a small core \citep{2009ApJ...696.2054G}.
On the other hand, currently large $Q'_{\rm p}$ are not necessarily at variance with 
the values required for the circularisation of very eccentric initial orbits as assumed by HEM 
($10^5 \la Q'_{\rm p} \la 5 \times 10^6$, cf. \citealt{2011Natur.473..187N, 2012arXiv1209.5724S}). 
This happens because $Q'_{\rm p}$ is a function of the tidal frequency and 
it is ill-defined in the case of highly eccentric orbits. 
In such a case, tidal dissipation is a highly non-sinusoidal function of the time strongly peaked around periastron 
so that an impulse approximation is much more adequate for its description (cf. \citealt{2008ApJ...678..498N, 2014ARA&A..52..171O}). 
Moreover, we lack observational constraints on the initial orbital and planetary parameters 
and on the duration of  the circularisation process because young hot Jupiters are very difficult 
to discover and confirm owing to the high level of stellar activity. 
This inevitably makes any estimate of the tidal dissipation rate during HEM uncertain.

Stellar modified tidal quality factors $Q'_{\rm s} \gtrsim 10^{6}-10^{7}$ were found  
for stars with eccentric planets at short orbital distances $a < 0.05$~au. This range is consistent with 
that estimated from tidal evolution calculations for a few individual systems  
such as OGLE-TR-56b \citep{2007P&SS...55..643C} and CoRoT-11b \citep{2011A&A...529A..50L},
and with the results by \citet{2012ApJ...757....6H}. 
Tidal evolution models for the population of hot Jupiters as computed by \citet{2008ApJ...678.1396J, 2009ApJ...698.1357J}
are also in general agreement with $10^6 \la Q'_{\rm s} \la 10^7$, 
but our ignorance of the initial conditions of the close giant planet population ultimately hampers 
any possibility of precisely estimating $Q'_{\rm s}$. We note that our values are also in general agreement 
with those theoretically expected in the case of stars hosting hot Jupiters according to the dynamic tide theory in 
\citet{2007ApJ...661.1180O} that may account for the differences in the estimated $Q'_{\rm s}$ 
between close stellar binary systems and star-planet systems.

Our catalogue will be updated at the end of our survey by including new HARPS-N data, additional
constraints from future secondary eclipse observations, and newly discovered giant planets, 
among which those announced in 2016, as these are detected by ground-based and space-based transit surveys. 
To this end, we stress the importance of collecting for each new system not only 
the RV data that are required for the confirmation of its planetary nature, but also -- when possible -- 
sufficient and precise enough RV measurements that permit an accurate determination of its orbital parameters and,
specifically, its eccentricity. Indeed, almost half of the TGPs in our sample have undetermined eccentricities.

We point out that different methods are currently used to determine stellar, hence planetary, parameters (mass, radius, and age)
such as stellar evolutionary tracks, empirical relations, or gyrochronology relations for stellar ages; 
sometimes, the last are then used as priors in stellar models. 
Although we used system ages estimated from evolutionary tracks 
(and recomputed them when only gyrochronologic estimates were available from the literature; see Sect.~\ref{system_parameters}), 
a homogeneous determination of stellar parameters may also be very useful to 
get a uniform catalogue of stellar and planetary parameters. 
However, this goes beyond the scope of the present work. 
In some cases, more accurate stellar parameters are also expected from absolute luminosities
as derived with Gaia parallaxes released in April 2018.

Long-term RV monitoring is also essential in order to discover long-period companions, 
derive their orbital parameters, and thus investigate their possible influence 
on the eccentricity and obliquity of the orbits of the close-in giant planets 
(e.g. \citealt{2010ApJ...725.1995M, 2014ApJ...785..126K}).
Future work will thus focus on detailed studies of the statistical properties and frequencies of close-in giant planets 
in connection with the presence of distant companions, 
taking advantage of the full temporal baseline ($\gtrsim 5-6$~years) 
of the HARPS-N observations upon conclusion of our survey.

\clearpage
\newpage

\bibliographystyle{aa} 
\bibliography{Bonomoetal_2017} 

\begin{acknowledgements}
We thank the two anonymous referees for their comments that allowed us to improve the present manuscript. 
We are grateful to C.~A.~Latham and D.~Latham for providing us with the corrected and 
new radial-velocity measurements of Qatar-2 that were gathered with the TRES spectrograph
on the 1.5~m Tillinghast Reflector at the Fred L. Whipple Observatory, Arizona.
GAPS acknowledges support from INAF (Italian National Institute of Astrophysics) through the `Progetti Premiali'
funding scheme of the Italian Ministry of Education, University, and Research. 
A.S.B., L.A., and L.M. acknowledge funding from the European Union Seventh Framework programme (FP7/2007-2013) 
under grant agreement No. 313014 (ETAEARTH).
G.S. acknowledges financial support from ``Accordo ASI--INAF'' n. 2013-016-R.0 July 9, 2013.
We acknowledge the use of e-infrastructure and support provided by IA2 (Italian Center for Astronomical Archives) of INAF - http://ia2.inaf.it.
\end{acknowledgements}

\clearpage
\newpage


\onecolumn

\centering
\begin{longtable}{l c c c c c}
\caption{Summary of the radial-velocity data used in the present work. The number of the HARPS-N 
RVs in the third column, $N_{\rm RV}$~H-N, only refers to the 45 systems that we monitored. $N_{\rm Dat}$ is the number of 
different RV datasets per object.}\\ 
\hline
Star 	& $N_{\rm RV}$ 	& $N_{\rm RV}$	& Duration 	& $N_{\rm Dat}$	& Ref.  	\\
	&  tot			&  H-N			& days    		& 		  		& 		\\
\hline	
\endfirsthead
\caption{continued.}\\
\hline 
Star 	& $N_{\rm RV}$ 	& $N_{\rm RV}$	& Duration 	& $N_{\rm Dat}$	& Ref.  	\\
	&  tot			&  H-N			& days    		& 		  		& 		\\
\hline
\endhead
\hline
\endfoot
CoRoT-1 & 9 & ~-~ & 197 & 1 & \refrv{2008A&A...482L..17B} \\ 
CoRoT-2 & 23 & ~-~ & 54 & 2 & \refrv{2008A&A...482L..21A} \\ 
CoRoT-3 & 22 & ~-~ & 334 & 3 & \refrv{2008A&A...491..889D} \\ 
CoRoT-4 & 20 & ~-~ & 102 & 2 & \refrv{2008A&A...488L..47M} \\ 
CoRoT-5 & 19 & ~-~ & 342 & 2 & \refrv{2009A&A...506..281R}  \\ 
CoRoT-6 & 14 & ~-~ & 73 & 1 & \refrv{2010A&A...512A..14F} \\ 
CoRoT-8 & 19 & ~-~ & 371 & 1 & \refrv{2010A&A...520A..66B} \\ 
CoRoT-9 & 28 & ~-~ & 1785 & 1 & \refrv{2010Natur.464..384D},\refrv{2017arXiv170306477B} \\    
CoRoT-10 & 19 & ~-~ & 438 & 1 & \refrv{2010A&A...520A..65B} \\ 
CoRoT-11 & 12 & ~-~ & 338 & 2 & \refrv{2010A&A...524A..55G} \\ 
CoRoT-12 & 20 & ~-~ & 488 & 3 & \refrv{2010A&A...520A..97G} \\ 
CoRoT-13 & 15 & ~-~ & 85 & 1 & \refrv{2010A&A...522A.110C} \\ 
CoRoT-14 & 14 & ~-~ & 90 & 2 & \refrv{2011A&A...528A..97T} \\ 
CoRoT-16 & 31 & ~-~ & 418 & 3 & \refrv{2012A&A...541A.149O} \\ 
CoRoT-17 & 17 & ~-~ & 101 & 1 & \refrv{2011A&A...531A..41C} \\ 
CoRoT-18 & 22 & ~-~ & 106 & 3 & \refrv{2011A&A...533A.130H} \\ 
CoRoT-19 & 22 & ~-~ & 114 & 3 & \refrv{2012A&A...537A.136G} \\ 
CoRoT-20 & 15 & ~-~ & 59 & 3 & \refrv{2012A&A...538A.145D} \\ 
CoRoT-21 & 19 & ~-~ & 455 & 2 & \refrv{2012A&A...545A...6P} \\ 
CoRoT-23 & 10 & ~-~ & 59 & 1 & \refrv{2012A&A...537A..54R} \\ 
CoRoT-25 & 34 & ~-~ & 814 & 2 & \refrv{2013A&A...555A.118A} \\ 
CoRoT-26 & 24 & ~-~ & 805 & 1 & 22 \\ 
CoRoT-27 & 13 & ~-~ & 69 & 1 & \refrv{2014A&A...562A.140P} \\ 
CoRoT-28 & 42 & ~-~ & 713 & 3 & \refrv{2015A&A...579A..36C} \\ 
CoRoT-29 & 20 & ~-~ & 417 & 1 & 24 \\   
HAT-P-1 & 61 & 11 & 3416 & 3 & \refrv{2007ApJ...656..552B},\refrv{2008ApJ...686..649J}  \\      
HAT-P-2 & 85 & 18 & 3226 & 3 & \refrv{2010MNRAS.401.2665P},\refrv{2008A&A...481..529L},\refrv{2007ApJ...670..826B},\refrv{2014ApJ...785..126K} \\ 
HAT-P-3 & 33 & 24 & 2927 & 2 & \refrv{2007ApJ...666L.121T}  \\ 
HAT-P-4 & 49 & 13 & 2722 & 3 & \refrv{2007ApJ...670L..41K},\refrv{2012MNRAS.422.3151H},30 \\      
HAT-P-5 & 8 & ~-~ & 28 & 1 & \refrv{2007ApJ...671L.173B}  \\ 
HAT-P-6 & 36 & 10 & 3251 & 2 & \refrv{2008ApJ...673L..79N},30\\ 
HAT-P-7 & 101 & 11 & 2873 & 4 & \refrv{2008ApJ...680.1450P},\refrv{2009ApJ...703L..99W},33,30\\     
HAT-P-8 & 43 & 12 & 2955 & 3 & \refrv{2009ApJ...704.1107L},\refrv{2011A&A...533A.113M},30 \\       
HAT-P-9 & 15 & ~-~ & 358 & 1 & \refrv{2009ApJ...690.1393S}  \\ 
HAT-P-12 & 23 & ~-~ & 2158 & 1 & \refrv{2009ApJ...706..785H},30 \\ 
HAT-P-13 & 66 & ~-~ & 2126 & 1 & \refrv{2009ApJ...707..446B},30 \\ 
HAT-P-14 & 59 & 11 & 2514 & 4 & \refrv{2010ApJ...715..458T},\refrv{2011AJ....141..161S},30 \\   
HAT-P-15 & 29 & ~-~ & 1859 & 1 & \refrv{2010ApJ...724..866K},30  \\ 
HAT-P-16 & 20 & 10 & 2256 & 2 & \refrv{2010ApJ...720.1118B},30  \\ 
HAT-P-17 & 72 & 25 & 2957 & 2 & \refrv{2012ApJ...749..134H},30 \\ 
HAT-P-18 & 48 & 17 & 2747 & 2 & \refrv{2011ApJ...726...52H},30  \\ 
HAT-P-19 & 39 & ~-~ & 200 & 2 & 48 \\  
HAT-P-20 & 42 & 27 & 2077 & 2 & \refrv{2011ApJ...742..116B},30 \\ 
HAT-P-21 & 39 & 24 & 2224 & 2 & 49 \\ 
HAT-P-22 & 58 & 39 & 2620 & 2 & 49 \\ 
HAT-P-23 & 13 & ~-~ & 556 & 1 & 49 \\ 
HAT-P-24 & 36 & 13 & 2185 & 2 & \refrv{2010ApJ...725.2017K},30  \\ 
HAT-P-25 & 8 & ~-~ & 63 & 1 & \refrv{2012ApJ...745...80Q}  \\ 
HAT-P-27 & 27 & ~-~ & 503 & 3 & \refrv{2012ApJ...760..139B},\refrv{2011PASP..123..555A}  \\       
HAT-P-28 & 17 & ~-~ & 145 & 2 & \refrv{2011ApJ...733..116B}  \\ 
HAT-P-29 & 37 & 25 & 1805 & 2 & 54,30 \\ 
HAT-P-30 & 52 & 13 & 1685 & 4 & \refrv{2011ApJ...735...24J},\refrv{2011AJ....142...86E},30  \\       
HAT-P-31 & 25 & 14 & 1925 & 2 & \refrv{2011AJ....142...95K},30  \\ 
HAT-P-32 & 30 & ~-~ & 1828 & 1 & \refrv{2011ApJ...742...59H},30 \\ 
HAT-P-33 & 26 & ~-~ & 1538 & 1 & 58,30\\ 
HAT-P-34 & 33 & ~-~ & 814 & 3 & \refrv{2012AJ....144...19B},30 \\ 
HAT-P-35 & 12 & ~-~ & 79 & 2 & 59 \\ 
HAT-P-36 & 12 & ~-~ & 34 & 1 & 59 \\ 
HAT-P-37 & 13 & ~-~ & 52 & 1 & 59 \\ 
HAT-P-38 & 14 & ~-~ & 3 & 1 & \refrv{2012PASJ...64...97S} \\ 
HAT-P-39 & 28 & ~-~ & 745 & 1 & \refrv{2012AJ....144..139H} \\ 
HAT-P-40 & 13 & ~-~ & 662 & 1 & 61 \\ 
HAT-P-41 & 22 & ~-~ & 439 & 2 & 61 \\ 
HAT-P-42 & 14 & ~-~ & 51 & 1 & \refrv{2013A&A...558A..86B} \\ 
HAT-P-43 & 6 & ~-~ & 7 & 1 & 62 \\ 
HAT-P-45 & 10 & ~-~ & 323 & 1 & \refrv{2014AJ....147..128H} \\ 
HAT-P-49 & 22 & ~-~ & 127 & 2 & \refrv{2014AJ....147...84B} \\ 
HAT-P-50 & 30 & ~-~ & 289 & 3 & \refrv{2015AJ....150..168H} \\ 
HAT-P-51 & 26 & ~-~ & 114 & 2 & 65 \\ 
HAT-P-52 & 7 & ~-~ & 527 & 1 & 65 \\ 
HAT-P-53 & 6 & ~-~ & 95 & 1 & 65 \\ 
HAT-P-54 & 17 & ~-~ & 410 & 2 & \refrv{2015AJ....149..149B} \\ 
HAT-P-55 & 16 & ~-~ & 104 & 2 & \refrv{2015PASP..127..851J} \\ 
HAT-P-56 & 18 & ~-~ & 44 & 1 & \refrv{2015AJ....150...85H} \\ 
HATS-1 & 33 & ~-~ & 334 & 3 & \refrv{2013AJ....145....5P}  \\ 
HATS-2 & 14 & ~-~ & 334 & 2 & \refrv{2013A&A...558A..55M}  \\ 
HATS-3 & 29 & ~-~ & 67 & 3 & \refrv{2013AJ....146..113B}  \\ 
HATS-4 & 41 & ~-~ & 204 & 4 & \refrv{2014AJ....148...29J}  \\ 
HATS-5 & 21 & ~-~ & 164 & 2 & \refrv{2014AJ....147..144Z} \\ 
HATS-7 & 10 & ~-~ & 4 & 1 & \refrv{2015ApJ...813..111B} \\ 
HATS-8 & 9 & ~-~ & 85 & 1 & \refrv{2015AJ....150...49B} \\ 
HATS-9 & 22 & ~-~ & 258 & 3 & \refrv{2015AJ....150...33B} \\ 
HATS-10 & 21 & ~-~ & 331 & 3 & 76 \\ 
HATS-13 & 40 & ~-~ & 162 & 3 & \refrv{2015A&A...580A..63M} \\ 
HATS-14 & 17 & ~-~ & 543 & 2 & 77 \\        
HD\,149026 & 46 & ~-~ & 3325 & 2 & \refrv{2005ApJ...633..465S},\refrv{2006ApJ...646..505B},30 \\      
HD\,17156 & 97 & 54 & 3610 & 3 & \refrv{2007ApJ...669.1336F},\refrv{2009ApJ...693..794W} \\ 
HD\,189733 & 126 & ~-~ & 1504 & 3 & \refrv{2005A&A...444L..15B },\refrv{2006ApJ...653L..69W},\refrv{2009A&A...495..959B} \\     
HD\,209458 & 251 & ~-~ & 2923 & 3 & \refrv{2004A&A...414..351N },\refrv{2006ApJ...646..505B} \\    
HD\,80606 & 139 & ~-~ & 3480 & 3 & \refrv{2001A&A...375L..27N },\refrv{2009ApJ...703.2091W},\refrv{2009A&A...498L...5M} \\          
KELT-2A & 17 & ~-~ & 93 & 1 &  \refrv{2012ApJ...756L..39B} \\ 
KELT-3 & 24 & ~-~ & 332 & 3 & \refrv{2013ApJ...773...64P} \\ 
KELT-6 & 107 & 45 & 1178 & 3 & \refrv{2014AJ....147...39C},\refrv{2015A&A...581L...6D} \\   
KELT-8 & 12 & ~-~ & 86 & 1 & \refrv{2015ApJ...810...30F} \\ 
Kepler-5 & 8 & ~-~ & 124 & 1 & \refrv{2010ApJ...713L.131K} \\ 
Kepler-6 & 9 & ~-~ & 58 & 1 & \refrv{2010ApJ...713L.136D} \\ 
Kepler-7 & 9 & ~-~ & 9 & 1 & \refrv{2010ApJ...713L.140L} \\ 
Kepler-8 & 16 & ~-~ & 152 & 1 & \refrv{2010ApJ...724.1108J}  \\ 
Kepler-12 & 16 & ~-~ & 747 & 1 & \refrv{2011ApJS..197....9F}  \\ 
Kepler-15 & 30 & ~-~ & 225 & 2 & \refrv{2011ApJS..197...13E}  \\ 
Kepler-17 & 21 & ~-~ & 275 & 2 & \refrv{2011ApJS..197...14D},\refrv{2012A&A...538A..96B} \\    
Kepler-39 & 13 & ~-~ & 47 & 1 & \refrv{2011A&A...533A..83B} \\ 
Kepler-40 & 11 & ~-~ & 61 & 1 & \refrv{2011A&A...528A..63S} \\ 
Kepler-41 & 12 & ~-~ & 126 & 1 & \refrv{2011A&A...536A..70S} \\ 
Kepler-43 & 22 & ~-~ & 425 & 2 & 102,\refrv{2014ApJ...795..151E} \\  
Kepler-44 & 7 & ~-~ & 13 & 1 & 102 \\ 
Kepler-45 & 14 & ~-~ & 93 & 1 & \refrv{2012AJ....143..111J}\\ 
Kepler-74 & 17 & ~-~ & 170 & 2 & \refrv{2013A&A...554A.114H} \\  
Kepler-75 & 9 & ~-~ & 32 & 1 & 108 \\ 
Kepler-76 & 18 & ~-~ & 131 & 2 & \refrv{2013ApJ...771...26F}  \\ 
Kepler-77 & 30 & ~-~ & 835 & 3 & \refrv{2013A&A...557A..74G},106\\ 
Kepler-412 & 6 & ~-~ & 245 & 1 & \refrv{2014A&A...564A..56D} \\ 
Kepler-419 & 20 & ~-~ & 506 & 1 & \refrv{2014ApJ...791...89D} \\  
Kepler-422 & 15 & ~-~ & 778 & 1 & 106 \\  
Kepler-423 & 27 & ~-~ & 1146 & 2 & \refrv{2015A&A...576A..11G},106\\     
Kepler-424 & 26 & ~-~ & 653 & 1 & 106 \\ 
Kepler-425 & 10 & ~-~ & 339 & 1 & \refrv{2014A&A...572A..93H} \\ 
Kepler-426 & 10 & ~-~ & 338 & 1 & 114 \\ 
Kepler-427 & 9 & ~-~ & 338 & 1 & 114 \\ 
Kepler-428 & 6 & ~-~ & 337 & 1 & 114 \\ 
Kepler-432 & 139 & ~-~ & 1292 & 3 & \refrv{2015ApJ...803...49Q},\refrv{2015A&A...573L...6O},\refrv{2015A&A...573L...5C} \\        
Kepler-433 & 12 & ~-~ & 161 & 1 & \refrv{2015A&A...575A..71A} \\ 
Kepler-434 & 11 & ~-~ & 333 & 1 & 118 \\ 
Kepler-435 & 12 & ~-~ & 787 & 1 & 118 \\ 
OGLE-TR-10 & 16 & ~-~ & 728 & 2 & \refrv{2005A&A...431.1105B},\refrv{2005ApJ...624..372K}  \\ 
OGLE-TR-56 & 11 & ~-~ & 384 & 1 &  \refrv{2004ApJ...609.1071T}  \\ 
OGLE-TR-111 & 8 & ~-~ & 7 & 1 &  \refrv{2005A&A...438.1123P}  \\ 
OGLE-TR-113 & 14 & ~-~ & 395 & 2 &  \refrv{2004ApJ...609L..37K},\refrv{2005A&A...438.1123P}   \\  
OGLE-TR-132 & 4 & ~-~ & 3 & 1 &  \refrv{2004A&A...421L..13B}  \\ 
OGLE-TR-182 & 24 & ~-~ & 418 & 1 &  \refrv{2008A&A...487..749P}  \\ 
OGLE-TR-211 & 19 & ~-~ & 416 & 1 &  \refrv{2008A&A...482..299U}  \\ 
Qatar-1 & 15 & 15 & 1002 & 2$^1$ &  \refrv{2013A&A...554A..28C} \\ 
Qatar-2 & 69 & ~-~ & 470 & 1 &  \refrv{2012ApJ...750...84B} \\ 
TrES-1 & 23 & 15 & 4311 & 2 & \refrv{2004ApJ...613L.153A} \\ 
TrES-2 & 40 & 12 & 3041 & 3 & 33,30 \\ 
TrES-3 & 17 & ~-~ & 1947 & 2 & \refrv{2009ApJ...691.1145S},30  \\ 
TrES-4 & 24 & 18 & 2778 & 2 & \refrv{2015A&A...575L..15S},30  \\ 
TrES-5 & 8 & ~-~ & 217 & 1 & \refrv{2011ApJ...741..114M}  \\ 
WASP-1 & 41 & 10 & 3291 & 4 &  \refrv{2007MNRAS.375..951C},\refrv{2011MNRAS.414.3023S},30 \\   
WASP-2 & 22 & ~-~ & 2198 & 3 &  \refrv{2011ApJ...741..114M},33,134,30 \\   
WASP-3 & 21 & ~-~ & 1861 & 2 & \refrv{2008MNRAS.385.1576P},30 \\ 
WASP-4 & 32 & ~-~ & 2172 & 3 & \refrv{2008ApJ...675L.113W},33,30\\   
WASP-5 & 22 & ~-~ & 414 & 2 & \refrv{2008MNRAS.387L...4A},33 \\         
WASP-6 & 59 & ~-~ & 396 & 2 &  \refrv{2009A&A...501..785G} \\     
WASP-7 & 40 & ~-~ & 1879 & 3 &  \refrv{2009ApJ...690L..89H},33,30\\    
WASP-8 & 63 & ~-~ & 2099 & 3 &  \refrv{2010A&A...517L...1Q},30 \\ 
WASP-10 & 23 & ~-~ & 2192 & 3 &  \refrv{2009MNRAS.392.1585C},30 \\   
WASP-11 & 58 & 32 & 2797 & 4 & \refrv{2009A&A...502..395W},\refrv{2015A&A...579A.136M},30 \\       
WASP-12 & 59 & 15 & 2180 & 3 & \refrv{2011MNRAS.413.2500H},30\\ 
WASP-13 & 28 & 17 & 2662 & 2 & \refrv{2009A&A...502..391S} \\ 
WASP-14 & 59 & 12 & 2685 & 5 & \refrv{2009MNRAS.392.1532J},146,30 \\    
WASP-15 & 21 & ~-~ & 133 & 1 & \refrv{2009AJ....137.4834W},30\\ 
WASP-16 & 84 & ~-~ & 1574 & 3 &  \refrv{2009ApJ...703..752L},\refrv{ 2012MNRAS.423.1503B},30 \\        
WASP-17 & 84 & ~-~ & 1850 & 3 & \refrv{2010ApJ...709..159A},30 \\  
WASP-18 & 53 & ~-~ & 1849 & 3 & \refrv{2009Natur.460.1098H},\refrv{2010A&A...524A..25T},30   \\       
WASP-19 & 59 & ~-~ & 660 & 2 & \refrv{2010ApJ...708..224H},\refrv{2011ApJ...730L..31H},30  \\            
WASP-20 & 89 & ~-~ & 1918 & 2 & \refrv{2015A&A...575A..61A}  \\   
WASP-21 & 47 & 8 & 2551 & 4 & \refrv{2010A&A...519A..98B}  \\   
WASP-22 & 54 & ~-~ & 1934 & 3 & \refrv{2010AJ....140.2007M},30  \\ 
WASP-23 & 38 & ~-~ & 585 & 1 & \refrv{2011A&A...531A..24T}  \\ 
WASP-24 & 74 & 11 & 2318 & 4 & \refrv{2010ApJ...720..337S},135,30 \\     
WASP-25 & 28 & ~-~ & 181 & 1 & \refrv{2011MNRAS.410.1631E}  \\ 
WASP-26 & 41 & 9 & 2341 & 3 & \refrv{2010A&A...520A..56S},\refrv{2011A&A...534A..16A}  \\    
WASP-28 & 42 & ~-~ & 1267 & 2 & 157  \\ 
WASP-29 & 13 & ~-~ & 97 & 1 & \refrv{2010ApJ...723L..60H}  \\ 
WASP-31 & 74 & 12 & 2192 & 3 & \refrv{2011A&A...531A..60A},151  \\   
WASP-32 & 46 & 9 & 2268 & 3 & \refrv{2010PASP..122.1465M},52  \\    
WASP-34 & 32 & ~-~ & 1472 & 2 & \refrv{2011A&A...526A.130S},30 \\    
WASP-35 & 19 & 10 & 2091 & 2 & 56 \\ 
WASP-36 & 19 & ~-~ & 306 & 1 & \refrv{2012AJ....143...81S} \\ 
WASP-37 & 13 & ~-~ & 72 & 2 & \refrv{2011AJ....141....8S} \\ 
WASP-38 & 67 & 16 & 1910 & 4 & 52,135,30\\ 
WASP-39 & 17 & ~-~ & 76 & 2 & \refrv{2011A&A...531A..40F}\\ 
WASP-41 & 22 & ~-~ & 214 & 1 & \refrv{2011PASP..123..547M} \\ 
WASP-42 & 53 & ~-~ & 401 & 2 & \refrv{2012A&A...544A..72L} \\  
WASP-43 & 42 & 27 & 1940 & 2 & \refrv{2011A&A...535L...7H} \\ 
WASP-44 & 15 & ~-~ & 104 & 1 & \refrv{2012MNRAS.422.1988A} \\ 
WASP-45 & 13 & ~-~ & 94 & 1 &  175 \\ 
WASP-46 & 16 & ~-~ & 120 & 1 &  175 \\ 
WASP-48 & 25 & 11 & 1954 & 2 &  56 \\    
WASP-49 & 32 & 7 & 2045 & 2 &  173 \\ 
WASP-50 & 25 & 10 & 1738 & 2 & \refrv{2011A&A...533A..88G} \\ 
WASP-52 & 41 & ~-~ & 445 & 3 & \refrv{2013A&A...549A.134H}  \\  
WASP-54 & 36 & 13 & 1657 & 3 & \refrv{2013A&A...551A..73F} \\ 
WASP-55 & 20 & ~-~ & 173 & 1 & \refrv{2012MNRAS.426..739H} \\ 
WASP-56 & 13 & ~-~ & 40 & 1 &  178 \\ 
WASP-57 & 25 & ~-~ & 140 & 2 &  178 \\ 
WASP-58 & 15 & ~-~ & 33 & 1 &  177 \\ 
WASP-59 & 12 & ~-~ & 32 & 1 &  177 \\ 
WASP-60 & 31 & ~-~ & 474 & 2 &  177 \\ 
WASP-61 & 15 & ~-~ & 298 & 1 &  179 \\ 
WASP-62 & 25 & ~-~ & 370 & 1 &  179 \\ 
WASP-63 & 23 & ~-~ & 423 & 1 &  179 \\ 
WASP-64 & 16 & ~-~ & 250 & 1 & \refrv{2013A&A...552A..82G} \\ 
WASP-65 & 18 & ~-~ & 321 & 1 & \refrv{2013A&A...559A..36G} \\ 
WASP-66 & 30 & ~-~ & 428 & 1 &  179 \\ 
WASP-67 & 19 & ~-~ & 86 & 1 &  179 \\ 
WASP-68 & 43 & ~-~ & 824 & 1 & \refrv{2014A&A...563A.143D} \\ 
WASP-69 & 22 & ~-~ & 736 & 1 & \refrv{2014MNRAS.445.1114A} \\ 
WASP-70A & 26 & ~-~ & 1072 & 2 & 183 \\ 
WASP-71 & 22 & ~-~ & 87 & 2 & \refrv{2013A&A...552A.120S} \\ 
WASP-72 & 18 & ~-~ & 354 & 1 &  180 \\ 
WASP-73 & 20 & ~-~ & 712 & 1 &  182 \\ 
WASP-74 & 20 & ~-~ & 417 & 1 & \refrv{2015AJ....150...18H} \\ 
WASP-75 & 15 & ~-~ & 350 & 1 &  181 \\ 
WASP-77A & 19 & ~-~ & 849 & 2 & \refrv{2013PASP..125...48M} \\ 
WASP-78 & 17 & ~-~ & 82 & 1 & \refrv{2012A&A...547A..61S} \\ 
WASP-79 & 21 & ~-~ & 422 & 1 &  187 \\ 
WASP-80 & 44 & ~-~ & 419 & 2 & \refrv{2013A&A...551A..80T} \\ 
WASP-83 & 28 & ~-~ & 710 & 1 &  185 \\ 
WASP-84 & 20 & ~-~ & 87 & 1 &  183 \\  
WASP-88 & 23 & ~-~ & 733 & 1 &  182 \\ 
WASP-89 & 20 & ~-~ & 734 & 1 &  185 \\
WASP-94A & 32 & ~-~ & 734 & 1 & \refrv{2014A&A...572A..49N} \\
WASP-95 & 14 & ~-~ & 297 & 1 & \refrv{2014MNRAS.440.1982H} \\ 
WASP-96 & 21 & ~-~ & 379 & 1 &  190 \\ 
WASP-97 & 12 & ~-~ & 64 & 1 &  190 \\ 
WASP-98 & 14 & ~-~ & 365 & 1 &  190 \\ 
WASP-99 & 20 & ~-~ & 356 & 1 &  190 \\ 
WASP-100 & 19 & ~-~ & 170 & 1 &  190 \\ 
WASP-101 & 19 & ~-~ & 796 & 1 &  190 \\ 
WASP-103 & 18 & ~-~ & 101 & 1 & \refrv{2014A&A...562L...3G} \\ 
WASP-104 & 21 & ~-~ & 169 & 2 & \refrv{2014A&A...570A..64S} \\ 
WASP-106 & 29 & ~-~ & 398 & 2 & 192 \\ 
WASP-117 & 95 & ~-~ & 681 & 2 & \refrv{2014A&A...568A..81L} \\ 
WTS-2 & 7 & ~-~ & 86 & 1 & \refrv{2014MNRAS.440.1470B} \\ 
XO-1 & 18 & 12 & 3338 & 2 & \refrv{2006ApJ...648.1228M}  \\ 
XO-2N & 72 & 32 & 2401 & 3 & \refrv{2007ApJ...671.2115B},\refrv{2015A&A...575A.111D},30\\ 
XO-3 & 75 & 19 & 2833 & 4 & \refrv{2008A&A...488..763H},\refrv{2011PASJ...63L..57H},30 \\    
XO-4 & 41 & 14 & 1699 & 3 & \refrv{2010PASJ...62L..61N},30 \\  
XO-5 & 34 & ~-~ & 2021 & 2 & \refrv{2008ApJ...686.1331B},\refrv{2009ApJ...700..783P},30 \\        
\hline
~\\
~\\
\label{table_rvdata}
\end{longtable}
\footnotemark[1]{The two datasets refer to HARPS-N data taken before and after the replacement of the 
HARPS-N CCD in September 2012 because of the failure of the CCD red side readout. 
Owing to different RV zero points, these data were considered as independent datasets.} \\
\tablebib{(\refrvbib{2008A&A...482L..17B})~\citealt{2008A&A...482L..17B}; 
(\refrvbib{2008A&A...482L..21A})~\citealt{2008A&A...482L..21A}; 
(\refrvbib{2008A&A...491..889D})~\citealt{2008A&A...491..889D}; 
(\refrvbib{2008A&A...488L..47M})~\citealt{2008A&A...488L..47M}; 
(\refrvbib{2009A&A...506..281R})~\citealt{2009A&A...506..281R}; 
(\refrvbib{2010A&A...512A..14F})~\citealt{2010A&A...512A..14F};
(\refrvbib{2010A&A...520A..66B})~\citealt{2010A&A...520A..66B}; 
(\refrvbib{2010Natur.464..384D})~\citealt{2010Natur.464..384D};
(\refrvbib{2017arXiv170306477B})~\citealt{2017arXiv170306477B}; 
(\refrvbib{2010A&A...520A..65B})~\citealt{2010A&A...520A..65B}; 
(\refrvbib{2010A&A...524A..55G})~\citealt{2010A&A...524A..55G};
(\refrvbib{2010A&A...520A..97G})~\citealt{2010A&A...520A..97G};
(\refrvbib{2010A&A...522A.110C})~\citealt{2010A&A...522A.110C};
(\refrvbib{2011A&A...528A..97T})~\citealt{2011A&A...528A..97T};
(\refrvbib{2012A&A...541A.149O})~\citealt{2012A&A...541A.149O}; 
(\refrvbib{2011A&A...531A..41C})~\citealt{2011A&A...531A..41C};
(\refrvbib{2011A&A...533A.130H})~\citealt{2011A&A...533A.130H};
(\refrvbib{2012A&A...537A.136G})~\citealt{2012A&A...537A.136G};
(\refrvbib{2012A&A...538A.145D})~\citealt{2012A&A...538A.145D};
(\refrvbib{2012A&A...545A...6P})~\citealt{2012A&A...545A...6P};
(\refrvbib{2012A&A...537A..54R})~\citealt{2012A&A...537A..54R};
(\refrvbib{2013A&A...555A.118A})~\citealt{2013A&A...555A.118A};
(\refrvbib{2014A&A...562A.140P})~\citealt{2014A&A...562A.140P};
(\refrvbib{2015A&A...579A..36C})~\citealt{2015A&A...579A..36C}; 
(\refrvbib{2007ApJ...656..552B})~\citealt{2007ApJ...656..552B};
(\refrvbib{2008ApJ...686..649J})~\citealt{2008ApJ...686..649J};
(\refrvbib{2010MNRAS.401.2665P})~\citealt{2010MNRAS.401.2665P};
(\refrvbib{2008A&A...481..529L})~\citealt{2008A&A...481..529L};
(\refrvbib{2007ApJ...670..826B})~\citealt{2007ApJ...670..826B};
(\refrvbib{2014ApJ...785..126K})~\citealt{2014ApJ...785..126K};  
(\refrvbib{2007ApJ...666L.121T})~\citealt{2007ApJ...666L.121T};
(\refrvbib{2007ApJ...670L..41K})~\citealt{2007ApJ...670L..41K};
(\refrvbib{2012MNRAS.422.3151H})~\citealt{2012MNRAS.422.3151H};
(\refrvbib{2007ApJ...671L.173B})~\citealt{2007ApJ...671L.173B};       
(\refrvbib{2008ApJ...673L..79N})~\citealt{2008ApJ...673L..79N};
(\refrvbib{2008ApJ...680.1450P})~\citealt{2008ApJ...680.1450P};
(\refrvbib{2009ApJ...703L..99W})~\citealt{2009ApJ...703L..99W};
(\refrvbib{2009ApJ...704.1107L})~\citealt{2009ApJ...704.1107L};
(\refrvbib{2011A&A...533A.113M})~\citealt{2011A&A...533A.113M};
(\refrvbib{2009ApJ...690.1393S})~\citealt{2009ApJ...690.1393S};
(\refrvbib{2009ApJ...706..785H})~\citealt{2009ApJ...706..785H};
(\refrvbib{2009ApJ...707..446B})~\citealt{2009ApJ...707..446B};
(\refrvbib{2010ApJ...715..458T})~\citealt{2010ApJ...715..458T};
(\refrvbib{2011AJ....141..161S})~\citealt{2011AJ....141..161S};
(\refrvbib{2010ApJ...724..866K})~\citealt{2010ApJ...724..866K}; 
(\refrvbib{2010ApJ...720.1118B})~\citealt{2010ApJ...720.1118B};
(\refrvbib{2012ApJ...749..134H})~\citealt{2012ApJ...749..134H};
(\refrvbib{2011ApJ...726...52H})~\citealt{2011ApJ...726...52H};
(\refrvbib{2011ApJ...742..116B})~\citealt{2011ApJ...742..116B}; 
(\refrvbib{2010ApJ...725.2017K})~\citealt{2010ApJ...725.2017K};
(\refrvbib{2012ApJ...745...80Q})~\citealt{2012ApJ...745...80Q};
(\refrvbib{2012ApJ...760..139B})~\citealt{2012ApJ...760..139B};
(\refrvbib{2011PASP..123..555A})~\citealt{2011PASP..123..555A};
(\refrvbib{2011ApJ...733..116B})~\citealt{2011ApJ...733..116B}; 
(\refrvbib{2011ApJ...735...24J})~\citealt{2011ApJ...735...24J};
(\refrvbib{2011AJ....142...86E})~\citealt{2011AJ....142...86E};
(\refrvbib{2011AJ....142...95K})~\citealt{2011AJ....142...95K};
(\refrvbib{2011ApJ...742...59H})~\citealt{2011ApJ...742...59H}; 
(\refrvbib{2012AJ....144...19B})~\citealt{2012AJ....144...19B};
(\refrvbib{2012PASJ...64...97S})~\citealt{2012PASJ...64...97S};
(\refrvbib{2012AJ....144..139H})~\citealt{2012AJ....144..139H};
(\refrvbib{2013A&A...558A..86B})~\citealt{2013A&A...558A..86B}; 
(\refrvbib{2014AJ....147..128H})~\citealt{2014AJ....147..128H};
(\refrvbib{2014AJ....147...84B})~\citealt{2014AJ....147...84B};
(\refrvbib{2015AJ....150..168H})~\citealt{2015AJ....150..168H};
(\refrvbib{2015AJ....149..149B})~\citealt{2015AJ....149..149B};
(\refrvbib{2015PASP..127..851J})~\citealt{2015PASP..127..851J};
(\refrvbib{2015AJ....150...85H})~\citealt{2015AJ....150...85H}; 
(\refrvbib{2013AJ....145....5P})~\citealt{2013AJ....145....5P};
(\refrvbib{2013A&A...558A..55M})~\citealt{2013A&A...558A..55M};
(\refrvbib{2013AJ....146..113B})~\citealt{2013AJ....146..113B};
(\refrvbib{2014AJ....148...29J})~\citealt{2014AJ....148...29J};
(\refrvbib{2014AJ....147..144Z})~\citealt{2014AJ....147..144Z};
(\refrvbib{2015ApJ...813..111B})~\citealt{2015ApJ...813..111B};
(\refrvbib{2015AJ....150...49B})~\citealt{2015AJ....150...49B}; 
(\refrvbib{2015AJ....150...33B})~\citealt{2015AJ....150...33B}; 
(\refrvbib{2015A&A...580A..63M})~\citealt{2015A&A...580A..63M};
(\refrvbib{2005ApJ...633..465S})~\citealt{2005ApJ...633..465S};
(\refrvbib{2006ApJ...646..505B})~\citealt{2006ApJ...646..505B}; 
(\refrvbib{2007ApJ...669.1336F})~\citealt{2007ApJ...669.1336F};    
(\refrvbib{2009ApJ...693..794W})~\citealt{2009ApJ...693..794W};    
(\refrvbib{2005A&A...444L..15B})~\citealt{2005A&A...444L..15B}; 
(\refrvbib{2006ApJ...653L..69W})~\citealt{2006ApJ...653L..69W}; 
(\refrvbib{2009A&A...495..959B})~\citealt{2009A&A...495..959B}; 
(\refrvbib{2004A&A...414..351N})~\citealt{2004A&A...414..351N};  
(\refrvbib{2006ApJ...646..505B})~\citealt{2006ApJ...646..505B}; 
(\refrvbib{2001A&A...375L..27N})~\citealt{2001A&A...375L..27N}; 
(\refrvbib{2009ApJ...703.2091W})~\citealt{2009ApJ...703.2091W}; 
(\refrvbib{2009A&A...498L...5M})~\citealt{2009A&A...498L...5M}; 
(\refrvbib{2012ApJ...756L..39B})~\citealt{2012ApJ...756L..39B}; 
(\refrvbib{2013ApJ...773...64P})~\citealt{2013ApJ...773...64P}; 
(\refrvbib{2014AJ....147...39C})~\citealt{2014AJ....147...39C}; 
(\refrvbib{2015A&A...581L...6D})~\citealt{2015A&A...581L...6D}; 
(\refrvbib{2015ApJ...810...30F})~\citealt{2015ApJ...810...30F}; 
(\refrvbib{2010ApJ...713L.131K})~\citealt{2010ApJ...713L.131K}; 
(\refrvbib{2010ApJ...713L.136D})~\citealt{2010ApJ...713L.136D}; 
(\refrvbib{2010ApJ...713L.140L})~\citealt{2010ApJ...713L.140L}; 
(\refrvbib{2010ApJ...724.1108J})~\citealt{2010ApJ...724.1108J}; 
(\refrvbib{2011ApJS..197....9F})~\citealt{2011ApJS..197....9F}; 
(\refrvbib{2011ApJS..197...13E})~\citealt{2011ApJS..197...13E}; 
(\refrvbib{2011ApJS..197...14D})~\citealt{2011ApJS..197...14D}; 
(\refrvbib{2012A&A...538A..96B})~\citealt{2012A&A...538A..96B}; 
(\refrvbib{2011A&A...533A..83B})~\citealt{2011A&A...533A..83B}; 
(\refrvbib{2011A&A...528A..63S})~\citealt{2011A&A...528A..63S}; 
(\refrvbib{2011A&A...536A..70S})~\citealt{2011A&A...536A..70S}; 
(\refrvbib{2014ApJ...795..151E})~\citealt{2014ApJ...795..151E}; 
(\refrvbib{2012AJ....143..111J})~\citealt{2012AJ....143..111J}; 
(\refrvbib{2013A&A...554A.114H})~\citealt{2013A&A...554A.114H}; 
(\refrvbib{2013ApJ...771...26F})~\citealt{2013ApJ...771...26F}; 
(\refrvbib{2013A&A...557A..74G})~\citealt{2013A&A...557A..74G}; 
(\refrvbib{2014A&A...564A..56D})~\citealt{2014A&A...564A..56D}; 
(\refrvbib{2014ApJ...791...89D})~\citealt{2014ApJ...791...89D}; 
(\refrvbib{2015A&A...576A..11G})~\citealt{2015A&A...576A..11G}; 
(\refrvbib{2014A&A...572A..93H})~\citealt{2014A&A...572A..93H}; 
(\refrvbib{2015ApJ...803...49Q})~\citealt{2015ApJ...803...49Q}; 
(\refrvbib{2015A&A...573L...6O})~\citealt{2015A&A...573L...6O}; 
(\refrvbib{2015A&A...573L...5C})~\citealt{2015A&A...573L...5C}; 
(\refrvbib{2015A&A...575A..71A})~\citealt{2015A&A...575A..71A}; 
(\refrvbib{2005A&A...431.1105B})~\citealt{2005A&A...431.1105B}; 
(\refrvbib{2005ApJ...624..372K})~\citealt{2005ApJ...624..372K}; 
(\refrvbib{2004ApJ...609.1071T})~\citealt{2004ApJ...609.1071T}; 
(\refrvbib{2005A&A...438.1123P})~\citealt{2005A&A...438.1123P}; 
(\refrvbib{2004ApJ...609L..37K})~\citealt{2004ApJ...609L..37K}; 
(\refrvbib{2005A&A...438.1123P})~\citealt{2005A&A...438.1123P}; 
(\refrvbib{2004A&A...421L..13B})~\citealt{2004A&A...421L..13B}; 
(\refrvbib{2008A&A...487..749P})~\citealt{2008A&A...487..749P}; 
(\refrvbib{2008A&A...482..299U})~\citealt{2008A&A...482..299U}; 
(\refrvbib{2013A&A...554A..28C})~\citealt{2013A&A...554A..28C}; 
(\refrvbib{2012ApJ...750...84B})~\citealt{2012ApJ...750...84B}; 
(\refrvbib{2004ApJ...613L.153A})~\citealt{2004ApJ...613L.153A}; 
(\refrvbib{2009ApJ...691.1145S})~\citealt{2009ApJ...691.1145S}; 
(\refrvbib{2015A&A...575L..15S})~\citealt{2015A&A...575L..15S}; 
(\refrvbib{2011ApJ...741..114M})~\citealt{2011ApJ...741..114M}; 
(\refrvbib{2007MNRAS.375..951C})~\citealt{2007MNRAS.375..951C}; 
(\refrvbib{2011MNRAS.414.3023S})~\citealt{2011MNRAS.414.3023S}; 
(\refrvbib{2011ApJ...741..114M})~\citealt{2011ApJ...741..114M}; 
(\refrvbib{2008MNRAS.385.1576P})~\citealt{2008MNRAS.385.1576P}; 
(\refrvbib{2008ApJ...675L.113W})~\citealt{2008ApJ...675L.113W}; 
(\refrvbib{2008MNRAS.387L...4A})~\citealt{2008MNRAS.387L...4A}; 
(\refrvbib{2009A&A...501..785G})~\citealt{2009A&A...501..785G}; 
(\refrvbib{2009ApJ...690L..89H})~\citealt{2009ApJ...690L..89H}; 
(\refrvbib{2010A&A...517L...1Q})~\citealt{2010A&A...517L...1Q}; 
(\refrvbib{2009MNRAS.392.1585C})~\citealt{2009MNRAS.392.1585C}; 
(\refrvbib{2009A&A...502..395W})~\citealt{2009A&A...502..395W}; 
(\refrvbib{2015A&A...579A.136M})~\citealt{2015A&A...579A.136M}; 
(\refrvbib{2011MNRAS.413.2500H})~\citealt{2011MNRAS.413.2500H}; 
(\refrvbib{2009A&A...502..391S})~\citealt{2009A&A...502..391S}; 
(\refrvbib{2009MNRAS.392.1532J})~\citealt{2009MNRAS.392.1532J}; 
(\refrvbib{2009AJ....137.4834W})~\citealt{2009AJ....137.4834W}; 
(\refrvbib{2009ApJ...703..752L})~\citealt{2009ApJ...703..752L}; 
(\refrvbib{2012MNRAS.423.1503B})~\citealt{2012MNRAS.423.1503B}; 
(\refrvbib{2010ApJ...709..159A})~\citealt{2010ApJ...709..159A}; 
(\refrvbib{2009Natur.460.1098H})~\citealt{2009Natur.460.1098H}; 
(\refrvbib{2010A&A...524A..25T})~\citealt{2010A&A...524A..25T}; 
(\refrvbib{2010ApJ...708..224H})~\citealt{2010ApJ...708..224H}; 
(\refrvbib{2011ApJ...730L..31H})~\citealt{2011ApJ...730L..31H}; 
(\refrvbib{2015A&A...575A..61A})~\citealt{2015A&A...575A..61A}; 
(\refrvbib{2010A&A...519A..98B})~\citealt{2010A&A...519A..98B}; 
(\refrvbib{2010AJ....140.2007M})~\citealt{2010AJ....140.2007M}; 
(\refrvbib{2011A&A...531A..24T})~\citealt{2011A&A...531A..24T}; 
(\refrvbib{2010ApJ...720..337S})~\citealt{2010ApJ...720..337S}; 
(\refrvbib{2011MNRAS.410.1631E})~\citealt{2011MNRAS.410.1631E}; 
(\refrvbib{2010A&A...520A..56S})~\citealt{2010A&A...520A..56S}; 
(\refrvbib{2011A&A...534A..16A})~\citealt{2011A&A...534A..16A}; 
(\refrvbib{2010ApJ...723L..60H})~\citealt{2010ApJ...723L..60H}; 
(\refrvbib{2011A&A...531A..60A})~\citealt{2011A&A...531A..60A}; 
(\refrvbib{2010PASP..122.1465M})~\citealt{2010PASP..122.1465M}; 
(\refrvbib{2011A&A...526A.130S})~\citealt{2011A&A...526A.130S};   
(\refrvbib{2012AJ....143...81S})~\citealt{2012AJ....143...81S}; 
(\refrvbib{2011AJ....141....8S})~\citealt{2011AJ....141....8S}; 
(\refrvbib{2011A&A...531A..40F})~\citealt{2011A&A...531A..40F}; 
(\refrvbib{2011PASP..123..547M})~\citealt{2011PASP..123..547M}; 
(\refrvbib{2012A&A...544A..72L})~\citealt{2012A&A...544A..72L}; 
(\refrvbib{2011A&A...535L...7H})~\citealt{2011A&A...535L...7H}; 
(\refrvbib{2012MNRAS.422.1988A})~\citealt{2012MNRAS.422.1988A}; 
(\refrvbib{2011A&A...533A..88G})~\citealt{2011A&A...533A..88G}; 
(\refrvbib{2013A&A...549A.134H})~\citealt{2013A&A...549A.134H}; 
(\refrvbib{2013A&A...551A..73F})~\citealt{2013A&A...551A..73F}; 
(\refrvbib{2012MNRAS.426..739H})~\citealt{2012MNRAS.426..739H}; 
(\refrvbib{2013A&A...552A..82G})~\citealt{2013A&A...552A..82G}; 
(\refrvbib{2013A&A...559A..36G})~\citealt{2013A&A...559A..36G}; 
(\refrvbib{2014A&A...563A.143D})~\citealt{2014A&A...563A.143D}; 
(\refrvbib{2014MNRAS.445.1114A})~\citealt{2014MNRAS.445.1114A}; 
(\refrvbib{2013A&A...552A.120S})~\citealt{2013A&A...552A.120S}; 
(\refrvbib{2015AJ....150...18H})~\citealt{2015AJ....150...18H}; 
(\refrvbib{2013PASP..125...48M})~\citealt{2013PASP..125...48M}; 
(\refrvbib{2012A&A...547A..61S})~\citealt{2012A&A...547A..61S}; 
(\refrvbib{2013A&A...551A..80T})~\citealt{2013A&A...551A..80T}; 
(\refrvbib{2014A&A...572A..49N})~\citealt{2014A&A...572A..49N}; 
(\refrvbib{2014MNRAS.440.1982H})~\citealt{2014MNRAS.440.1982H}; 
(\refrvbib{2014A&A...562L...3G})~\citealt{2014A&A...562L...3G}; 
(\refrvbib{2014A&A...570A..64S})~\citealt{2014A&A...570A..64S}; 
(\refrvbib{2014A&A...568A..81L})~\citealt{2014A&A...568A..81L}; 
(\refrvbib{2014MNRAS.440.1470B})~\citealt{2014MNRAS.440.1470B}; 
(\refrvbib{2006ApJ...648.1228M})~\citealt{2006ApJ...648.1228M};
(\refrvbib{2007ApJ...671.2115B})~\citealt{2007ApJ...671.2115B};
(\refrvbib{2015A&A...575A.111D})~\citealt{2015A&A...575A.111D};
(\refrvbib{2008A&A...488..763H})~\citealt{2008A&A...488..763H};
(\refrvbib{2011PASJ...63L..57H})~\citealt{2011PASJ...63L..57H};
(\refrvbib{2010PASJ...62L..61N})~\citealt{2010PASJ...62L..61N};
(\refrvbib{2008ApJ...686.1331B})~\citealt{2008ApJ...686.1331B};
(\refrvbib{2009ApJ...700..783P})~\citealt{2009ApJ...700..783P};
}


\clearpage
\newpage



\centering
\begin{longtable}{l c c c c c}
\caption{Priors on orbital periods, and transit and occultation midtimes from ground-based and/or space-based photometry.}\\
\hline 
Planet & Period & Transit Midtime    				& Ephemeris 	& Occultation Midtime 				& Occultation   \\
           & days    &  $\rm BJD_{\rm TDB}-2,450,000$   	& Reference  	& $\rm BJD_{\rm TDB}-2,450,000$		& Reference     \\
\hline
\endfirsthead
\caption{continued.}\\
\hline 
Planet & Period & Transit Midtime    				& Ephemeris 	& Occultation Midtime 				& Occultation   \\
           & days    &  $\rm BJD_{\rm TDB}-2,450,000$   	& Reference  	& $\rm BJD_{\rm TDB}-2,450,000$		& Reference     \\
\hline
\endhead
\hline
\endfoot
CoRoT-1b & $1.50896820\pm\rm5.0\mbox{\sc{e}-07}$ & $4138.328070\pm\rm6.0\mbox{\sc{e}-05}$ & \refeph{2012PASP..124..212S} & $5162.1651\pm\rm0.0036$ & \refeph{2011ApJ...726...95D} \\ 
 &  &  &  & $5159.1441\pm\rm0.0021$ & 2 \\ 
CoRoT-2b & $1.74299673\pm\rm3.1\mbox{\sc{e}-07}$ & $5628.44758\pm\rm0.00014$ & \refeph{2015MNRAS.450.3101B} & $5160.4504\pm\rm0.0012$ & 2 \\ 
 &  &  &  & $4771.76060\pm\rm0.00070$ &  2 \\ 
 &  &  &  & $4771.7641\pm\rm0.0019$ & 2 \\ 
CoRoT-3b & $4.2567994\pm\rm3.5\mbox{\sc{e}-06}$ & $4283.13388\pm\rm0.00024$ & \refeph{2009A&A...506..377T} &  &  \\ 
CoRoT-4b & $9.20205\pm\rm0.00037$ & $4141.36493\pm\rm0.00089$ & \refeph{2008A&A...488L..43A} &  &  \\ 
CoRoT-5b & $4.0378962\pm\rm1.9\mbox{\sc{e}-06}$ & $4400.199630\pm\rm2.0\mbox{\sc{e}-05}$ & \refeph{2009A&A...506..281R} &  &  \\
\end{longtable}
\tablebib{(\refephbib{2012PASP..124..212S})~\citealt{2012PASP..124..212S}; 
(\refephbib{2011ApJ...726...95D})~\citealt{2011ApJ...726...95D}; 
(\refephbib{2015MNRAS.450.3101B})~\citealt{2015MNRAS.450.3101B}; 
(\refephbib{2009A&A...506..377T})~\citealt{2009A&A...506..377T}; 
(\refephbib{2008A&A...488L..43A})~\citealt{2008A&A...488L..43A}; 
(\refephbib{2009A&A...506..281R})~\citealt{2009A&A...506..281R};
(\refephbib{2010A&A...512A..14F})~\citealt{2010A&A...512A..14F};
(\refephbib{2011MNRAS.417.2166S})~\citealt{2011MNRAS.417.2166S};
(\refephbib{2017arXiv170306477B})~\citealt{2017arXiv170306477B};
(\refephbib{2010A&A...520A..65B})~\citealt{2010A&A...520A..65B};
(\refephbib{2012A&A...543L...5G})~\citealt{2012A&A...543L...5G};
(\refephbib{2010A&A...520A..97G})~\citealt{2010A&A...520A..97G};
(\refephbib{2010A&A...522A.110C})~\citealt{2010A&A...522A.110C};
(\refephbib{2011A&A...528A..97T})~\citealt{2011A&A...528A..97T};
(\refephbib{2012A&A...541A.149O})~\citealt{2012A&A...541A.149O};
(\refephbib{2011A&A...531A..41C})~\citealt{2011A&A...531A..41C};
(\refephbib{2011A&A...533A.130H})~\citealt{2011A&A...533A.130H};
(\refephbib{2012A&A...537A.136G})~\citealt{2012A&A...537A.136G};
(\refephbib{2012A&A...538A.145D})~\citealt{2012A&A...538A.145D};
(\refephbib{2012A&A...545A...6P})~\citealt{2012A&A...545A...6P};
(\refephbib{2012A&A...537A..54R})~\citealt{2012A&A...537A..54R};
(\refephbib{2013A&A...555A.118A})~\citealt{2013A&A...555A.118A};
(\refephbib{2014A&A...562A.140P})~\citealt{2014A&A...562A.140P};
(\refephbib{2015A&A...579A..36C})~\citealt{2015A&A...579A..36C};
(\refephbib{2014MNRAS.437...46N})~\citealt{2014MNRAS.437...46N};
(\refephbib{2010ApJ...708..498T})~\citealt{2010ApJ...708..498T};
(\refephbib{2013ApJ...766...95L})~\citealt{2013ApJ...766...95L};
(\refephbib{2011AJ....141..179C})~\citealt{2011AJ....141..179C};
(\refephbib{2013ApJ...770..102T})~\citealt{2013ApJ...770..102T};
(\refephbib{2012MNRAS.422.3099S})~\citealt{2012MNRAS.422.3099S};
(\refephbib{2008ApJ...673L..79N})~\citealt{2008ApJ...673L..79N};
(\refephbib{2012ApJ...746..111T})~\citealt{2012ApJ...746..111T};
(\refephbib{2013ApJ...764L..22M})~\citealt{2013ApJ...764L..22M};
(\refephbib{2010ApJ...710...97C})~\citealt{2010ApJ...710...97C};
(\refephbib{2013A&A...551A..11M})~\citealt{2013A&A...551A..11M};
(\refephbib{2012NewA...17..438D})~\citealt{2012NewA...17..438D};
(\refephbib{2012MNRAS.420.2580S})~\citealt{2012MNRAS.420.2580S};
(\refephbib{2011A&A...527A..85N})~\citealt{2011A&A...527A..85N};
(\refephbib{2010ApJ...724..866K})~\citealt{2010ApJ...724..866K};
(\refephbib{2013A&A...557A..30C})~\citealt{2013A&A...557A..30C};
(\refephbib{2012ApJ...749..134H})~\citealt{2012ApJ...749..134H};
(\refephbib{2011ApJ...726...52H})~\citealt{2011ApJ...726...52H};
(\refephbib{2015ApJ...810..118K})~\citealt{2015ApJ...810..118K};
(\refephbib{2014AN....335..797G})~\citealt{2014AN....335..797G};
(\refephbib{2015ApJ...805..132D})~\citealt{2015ApJ...805..132D};
(\refephbib{2011ApJ...742..116B})~\citealt{2011ApJ...742..116B};
(\refephbib{2013RMxAA..49...71R})~\citealt{2013RMxAA..49...71R};
(\refephbib{2014ApJ...781..109O})~\citealt{2014ApJ...781..109O};
(\refephbib{2013RAA....13..593W})~\citealt{2013RAA....13..593W};
(\refephbib{2012ApJ...745...80Q})~\citealt{2012ApJ...745...80Q};       
(\refephbib{2011ApJ...733..116B})~\citealt{2011ApJ...733..116B};
(\refephbib{Espositoetal_inprep})~Esposito et al., in prep.;
(\refephbib{2011ApJ...735...24J})~\citealt{2011ApJ...735...24J};
(\refephbib{2011AJ....142...95K})~\citealt{2011AJ....142...95K};
(\refephbib{2011ApJ...742...59H})~\citealt{2011ApJ...742...59H};
(\refephbib{2014ApJ...796..115Z})~\citealt{2014ApJ...796..115Z};
(\refephbib{2012AJ....144...19B})~\citealt{2012AJ....144...19B};
(\refephbib{2012PASJ...64...97S})~\citealt{2012PASJ...64...97S};
(\refephbib{2012AJ....144..139H})~\citealt{2012AJ....144..139H};
(\refephbib{2013A&A...558A..86B})~\citealt{2013A&A...558A..86B};
(\refephbib{2014AJ....147..128H})~\citealt{2014AJ....147..128H};
(\refephbib{2014AJ....147...84B})~\citealt{2014AJ....147...84B};
(\refephbib{2015AJ....150..168H})~\citealt{2015AJ....150..168H};
(\refephbib{2015AJ....149..149B})~\citealt{2015AJ....149..149B};
(\refephbib{2015PASP..127..851J})~\citealt{2015PASP..127..851J};
(\refephbib{2015AJ....150...85H})~\citealt{2015AJ....150...85H};
(\refephbib{2013AJ....145....5P})~\citealt{2013AJ....145....5P};
(\refephbib{2013A&A...558A..55M})~\citealt{2013A&A...558A..55M};
(\refephbib{2013AJ....146..113B})~\citealt{2013AJ....146..113B};
(\refephbib{2014AJ....148...29J})~\citealt{2014AJ....148...29J};
(\refephbib{2014AJ....147..144Z})~\citealt{2014AJ....147..144Z};
(\refephbib{2015ApJ...813..111B})~\citealt{2015ApJ...813..111B};
(\refephbib{2015AJ....150...49B})~\citealt{2015AJ....150...49B};
(\refephbib{2015AJ....150...33B})~\citealt{2015AJ....150...33B};
(\refephbib{2015A&A...580A..63M})~\citealt{2015A&A...580A..63M};
(\refephbib{2009ApJ...703..769K})~\citealt{2009ApJ...703..769K};
(\refephbib{2012ApJ...754..136S})~\citealt{2012ApJ...754..136S};
(\refephbib{2011ApJ...726....3N})~\citealt{2011ApJ...726....3N};
(\refephbib{2008ApJ...686.1341C})~\citealt{2008ApJ...686.1341C};
(\refephbib{2010ApJ...721.1861A})~\citealt{2010ApJ...721.1861A};
(\refephbib{2009ApJ...690..822K})~\citealt{2009ApJ...690..822K};
(\refephbib{2007ApJ...655..564K})~\citealt{2007ApJ...655..564K};
(\refephbib{2008ApJ...673..526K})~\citealt{2008ApJ...673..526K};
(\refephbib{2014ApJ...790...53Z})~\citealt{2014ApJ...790...53Z};
(\refephbib{2010A&A...516A..95H})~\citealt{2010A&A...516A..95H};
(\refephbib{2009Natur.457..562L})~\citealt{2009Natur.457..562L};
(\refephbib{2012ApJ...756L..39B})~\citealt{2012ApJ...756L..39B};
(\refephbib{2013ApJ...773...64P})~\citealt{2013ApJ...773...64P};
(\refephbib{2015A&A...581L...6D})~\citealt{2015A&A...581L...6D};
(\refephbib{2015ApJ...810...30F})~\citealt{2015ApJ...810...30F};
(\refephbib{2013A&A...560A.112M})~\citealt{2013A&A...560A.112M};
(\refephbib{2011ApJ...735L..12D})~\citealt{2011ApJ...735L..12D};
(\refephbib{2011ApJS..197....9F})~\citealt{2011ApJS..197....9F};
(\refephbib{2015A&A...575A..85B})~\citealt{2015A&A...575A..85B};
(\refephbib{2012MNRAS.426.1291S})~\citealt{2012MNRAS.426.1291S};
(\refephbib{2011A&A...536A..70S})~\citealt{2011A&A...536A..70S};
(\refephbib{2013ApJ...771...26F})~\citealt{2013ApJ...771...26F};
(\refephbib{2013A&A...557A..74G})~\citealt{2013A&A...557A..74G};
(\refephbib{2014A&A...564A..56D})~\citealt{2014A&A...564A..56D};
(\refephbib{2014ApJ...791...89D})~\citealt{2014ApJ...791...89D};
(\refephbib{2014ApJ...795..151E})~\citealt{2014ApJ...795..151E};
(\refephbib{2015A&A...576A..11G})~\citealt{2015A&A...576A..11G};
(\refephbib{2014A&A...572A..93H})~\citealt{2014A&A...572A..93H};
(\refephbib{2015ApJ...803...49Q})~\citealt{2015ApJ...803...49Q};
(\refephbib{2015A&A...575A..71A})~\citealt{2015A&A...575A..71A};
(\refephbib{2009A&A...505..901B})~\citealt{2009A&A...505..901B};
(\refephbib{2011ApJ...741..102A})~\citealt{2011ApJ...741..102A};
(\refephbib{2011ApJ...733...53H})~\citealt{2011ApJ...733...53H};
(\refephbib{2010ApJ...721.1829A})~\citealt{2010ApJ...721.1829A};
(\refephbib{2011ApJ...728..125A})~\citealt{2011ApJ...728..125A};    
(\refephbib{2008A&A...487..749P})~\citealt{2008A&A...487..749P};
(\refephbib{2008A&A...482..299U})~\citealt{2008A&A...482..299U};
(\refephbib{2013A&A...554A..28C})~\citealt{2013A&A...554A..28C};
(\refephbib{2015ApJ...802...28C})~\citealt{2015ApJ...802...28C};
(\refephbib{2014MNRAS.443.2391M})~\citealt{2014MNRAS.443.2391M};
(\refephbib{2005ApJ...626..523C})~\citealt{2005ApJ...626..523C};
(\refephbib{2014ApJ...797...42C})~\citealt{2014ApJ...797...42C};
(\refephbib{2014MNRAS.444.1351R})~\citealt{2014MNRAS.444.1351R};
(\refephbib{2010ApJ...710.1551O})~\citealt{2010ApJ...710.1551O};
(\refephbib{2010ApJ...717.1084C})~\citealt{2010ApJ...717.1084C}; 
(\refephbib{2013ApJ...764....8K})~\citealt{2013ApJ...764....8K};
(\refephbib{2010ApJ...711..374F})~\citealt{2010ApJ...711..374F};
(\refephbib{2010ApJ...718..920C})~\citealt{2010ApJ...718..920C};
(\refephbib{2015A&A...575L..15S})~\citealt{2015A&A...575L..15S};
(\refephbib{2009ApJ...691..866K})~\citealt{2009ApJ...691..866K};
(\refephbib{2011ApJ...741..114M})~\citealt{2011ApJ...741..114M};
(\refephbib{2010arXiv1004.0836W})~\citealt{2010arXiv1004.0836W}; 
(\refephbib{2013AJ....146..147M})~\citealt{2013AJ....146..147M};
(\refephbib{2014MNRAS.441.3666R})~\citealt{2014MNRAS.441.3666R};
(\refephbib{2011A&A...530A...5C})~\citealt{2011A&A...530A...5C};
(\refephbib{2011ApJ...727...23B})~\citealt{2011ApJ...727...23B};
(\refephbib{2014A&A...564A...6C})~\citealt{2014A&A...564A...6C};
(\refephbib{2015MNRAS.447..463N})~\citealt{2015MNRAS.447..463N};
(\refephbib{2012ApJ...744..189A})~\citealt{2012ApJ...744..189A};
(\refephbib{2010A&A...517L...1Q})~\citealt{2010A&A...517L...1Q};
(\refephbib{2013ApJ...768...42C})~\citealt{2013ApJ...768...42C};
(\refephbib{2013MNRAS.430.3032B})~\citealt{2013MNRAS.430.3032B};
(\refephbib{2013A&A...551A.108M})~\citealt{2013A&A...551A.108M};
(\refephbib{2011ApJ...727..125C})~\citealt{2011ApJ...727..125C};
(\refephbib{2015ApJ...802...28C})~\citealt{2015ApJ...802...28C};
(\refephbib{2012MNRAS.419.1248B})~\citealt{2012MNRAS.419.1248B};
(\refephbib{2013ApJ...779....5B})~\citealt{2013ApJ...779....5B};
(\refephbib{2015ApJ...811..122W})~\citealt{2015ApJ...811..122W};
(\refephbib{2013MNRAS.434.1300S})~\citealt{2013MNRAS.434.1300S};
(\refephbib{2012MNRAS.426.1338S})~\citealt{2012MNRAS.426.1338S};
(\refephbib{2011MNRAS.416.2108A})~\citealt{2011MNRAS.416.2108A};
(\refephbib{2013MNRAS.428.2645M})~\citealt{2013MNRAS.428.2645M};
(\refephbib{2011ApJ...742...35N})~\citealt{2011ApJ...742...35N};
(\refephbib{2013MNRAS.436....2M})~\citealt{2013MNRAS.436....2M};
(\refephbib{2013ApJ...771..108B})~\citealt{2013ApJ...771..108B};
(\refephbib{2010MNRAS.404L.114G})~\citealt{2010MNRAS.404L.114G};
(\refephbib{2015A&A...575A..61A})~\citealt{2015A&A...575A..61A};
(\refephbib{2011A&A...534A..16A})~\citealt{2011A&A...534A..16A};
(\refephbib{2013A&A...553A..26N})~\citealt{2013A&A...553A..26N};
(\refephbib{2014MNRAS.444..776S})~\citealt{2014MNRAS.444..776S};
(\refephbib{2012A&A...545A..93S})~\citealt{2012A&A...545A..93S};
(\refephbib{2013MNRAS.432..693M})~\citealt{2013MNRAS.432..693M};
(\refephbib{2013MNRAS.428.3680G})~\citealt{2013MNRAS.428.3680G};
(\refephbib{2015MNRAS.446.2428S})~\citealt{2015MNRAS.446.2428S};
(\refephbib{2011A&A...526A.130S})~\citealt{2011A&A...526A.130S};
(\refephbib{2011AJ....142...86E})~\citealt{2011AJ....142...86E};
(\refephbib{2012AJ....143...81S})~\citealt{2012AJ....143...81S};
(\refephbib{2011AJ....141....8S})~\citealt{2011AJ....141....8S};
(\refephbib{2011A&A...525A..54B})~\citealt{2011A&A...525A..54B};
(\refephbib{2011A&A...531A..40F})~\citealt{2011A&A...531A..40F};
(\refephbib{2011PASP..123..547M})~\citealt{2011PASP..123..547M};
(\refephbib{2012A&A...544A..72L})~\citealt{2012A&A...544A..72L};
(\refephbib{2014A&A...563A..40C})~\citealt{2014A&A...563A..40C};
(\refephbib{2012A&A...542A...4G})~\citealt{2012A&A...542A...4G};
(\refephbib{2014ApJ...781..116B})~\citealt{2014ApJ...781..116B};
(\refephbib{2013MNRAS.430.2932M})~\citealt{2013MNRAS.430.2932M};
(\refephbib{2012MNRAS.422.1988A})~\citealt{2012MNRAS.422.1988A};
(\refephbib{2014A&A...567A...8C})~\citealt{2014A&A...567A...8C};
(\refephbib{2013MNRAS.431..966T})~\citealt{2013MNRAS.431..966T};
(\refephbib{2013A&A...551A..73F})~\citealt{2013A&A...551A..73F};
(\refephbib{2012MNRAS.426..739H})~\citealt{2012MNRAS.426..739H};
(\refephbib{2015MNRAS.454.3094S})~\citealt{2015MNRAS.454.3094S};
(\refephbib{2013A&A...549A.134H})~\citealt{2013A&A...549A.134H};
(\refephbib{2013A&A...552A..82G})~\citealt{2013A&A...552A..82G};
(\refephbib{2013A&A...559A..36G})~\citealt{2013A&A...559A..36G};
(\refephbib{2014A&A...563A.143D})~\citealt{2014A&A...563A.143D};
(\refephbib{2014MNRAS.445.1114A})~\citealt{2014MNRAS.445.1114A};
(\refephbib{2013A&A...552A.120S})~\citealt{2013A&A...552A.120S};
(\refephbib{2015AJ....150...18H})~\citealt{2015AJ....150...18H};
(\refephbib{2013PASP..125...48M})~\citealt{2013PASP..125...48M};
(\refephbib{2012A&A...547A..61S})~\citealt{2012A&A...547A..61S};
(\refephbib{2014ApJ...790..108F})~\citealt{2014ApJ...790..108F};
(\refephbib{2014A&A...572A..49N})~\citealt{2014A&A...572A..49N};
(\refephbib{2014MNRAS.440.1982H})~\citealt{2014MNRAS.440.1982H};
(\refephbib{2015MNRAS.447..711S})~\citealt{2015MNRAS.447..711S};
(\refephbib{2014A&A...570A..64S})~\citealt{2014A&A...570A..64S};
(\refephbib{2014A&A...568A..81L})~\citealt{2014A&A...568A..81L};
(\refephbib{2014MNRAS.440.1470B})~\citealt{2014MNRAS.440.1470B};
(\refephbib{2010ApJ...719.1796B})~\citealt{2010ApJ...719.1796B};
(\refephbib{2008ApJ...684.1427M})~\citealt{2008ApJ...684.1427M};
(\refephbib{2015A&A...575A.111D})~\citealt{2015A&A...575A.111D};
(\refephbib{2009ApJ...701..514M})~\citealt{2009ApJ...701..514M};
(\refephbib{2014ApJ...794..134W})~\citealt{2014ApJ...794..134W};
(\refephbib{2014arXiv1412.0451S})~\citealt{2014arXiv1412.0451S}.
}


\clearpage
\newpage


\tiny

\begin{landscape}
\centering
\begin{longtable}{l c c c c c c c c c c} 
\caption{System parameters that were used, along with the orbital parameters in Table~\ref{table_orbital_parameters}, to update
the mass, density, and surface gravity of the 231 transiting giant planets in our sample (see Table~\ref{table_planet_parameters}).
The vast majority of them are taken from the literature, although estimates of ages for 37 systems with stellar evolutionary tracks
are provided here for the first time (Sect.~\ref{system_parameters}). In a very few cases, namely CoRoT-1, WASP-63, WASP-78, and WASP-79, 
we adopted our values of stellar mass and stellar/planetary radii (see Sect.~\ref{system_parameters}).}\\
\hline 
System & $M_{\rm s}$ & $R_{\rm s}$ & $T_{\rm eff}$ & $\rm [Fe/H]$ & $\vsini$   & $P_{\rm rot}$ & Age   & $i_{\rm p}$   & $\rp$       & System    \\
             & [$\Msun$]     & [$\Rsun$]     & [K]               &   [dex]             & [$\kms$] & [days]             & [Gyr]  & [deg]           & [$\Rjup$] & Reference  \\
\hline
\endfirsthead
\caption{continued.}\\
\hline 
System & $M_{\rm s}$ & $R_{\rm s}$ & $T_{\rm eff}$ & $\rm [Fe/H]$ & $\vsini$   & $P_{\rm rot}$ & Age   & $i_{\rm p}$   & $\rp$       & System    \\
             & [$\Msun$]     & [$\Rsun$]     & [K]               &   dex       	   & [$\kms$] & [days]             & [Gyr]  & [deg]           & [$\Rjup$] & Reference \\
\hline
\endhead
\hline
\endfoot
%
%
CoRoT-1 & $1.220\pm0.030$ & $1.230\pm0.020$ & $6298\pm66$ & $0.060\pm0.070$ & $5.2\pm1.0$ & ~-~ & $1.60\pm0.50$ & $85.10\pm0.50$ & $1.715\pm0.030$ & \refsys{2008A&A...482L..17B}, this work$^1$\\ 
CoRoT-2 & $0.970\pm0.060$ & $0.902\pm0.018$ & $5625\pm120$ & $0.0\pm0.1$ & $11.85\pm0.50$ & $4.522\pm0.024$ & $2.7_{-2.7}^{+3.2}$ & $88.08_{-0.16}^{+0.18}$ & $1.466_{-0.044}^{+0.042}$ & \refsys{2008A&A...482L..25B},  \refsys{2008A&A...482L..21A},  \refsys{2010A&A...511A...3G}, \refsys{2009A&A...493..193L} \\ 
CoRoT-3 & $1.370\pm0.090$ & $1.560\pm0.090$ & $6740\pm140$ & $-0.020\pm0.060$ & $17.0\pm1.0$ & ~-~ & $2.20\pm0.60$ & $85.90\pm0.80$ & $1.010\pm0.070$ & \refsys{2008A&A...491..889D} \\ 
CoRoT-4 & $1.160_{-0.020}^{+0.030}$ & $1.1700_{-0.030}^{+0.0100}$ & $6190\pm60$ & $0.050\pm0.070$ & $6.4\pm1.0$ & $8.9\pm1.1$ & $1.00_{-0.30}^{+1.0}$ & $90.000_{-0.085}^{+0.00}$ & $1.190\pm0.055$ & \refsys{2008A&A...488L..47M} \\ 
CoRoT-5 & $1.000\pm0.020$ & $1.186\pm0.040$ & $6100\pm65$ & $-0.250\pm0.060$ & $1.0\pm1.0$ & ~-~ & $6.9\pm1.4$ & $85.83_{-1.4}^{+0.99}$ & $1.388_{-0.047}^{+0.046}$ & \refsys{2009A&A...506..281R} \\ 
CoRoT-6 & $1.050\pm0.050$ & $1.025\pm0.026$ & $6090\pm50$ & $-0.200\pm0.100$ & $7.5\pm1.0$ & $6.40\pm0.50$ & $2.5_{-1.7}^{+2.1}$ & $89.07\pm0.30$ & $1.166\pm0.035$ & \refsys{2010A&A...512A..14F}, \refsys{2011MNRAS.417.2166S} \\ 
CoRoT-8 & $0.880\pm0.040$ & $0.770\pm0.020$ & $5080\pm80$ & $0.300\pm0.100$ & $2.0\pm1.0$ & ~-~ & $3.00_{-3.0}^{+0.00}$ & $88.400\pm0.100$ & $0.570\pm0.020$ & \refsys{2010A&A...520A..66B} \\ 
CoRoT-9 & $0.960\pm0.040$ & $0.960\pm0.060$ & $5625\pm80$ & $-0.010\pm0.060$ & $3.50_{-3.5}^{+0.00}$ & ~-~ & $6.0\pm3.0$ & 
$89.900_{-0.084}^{+0.066}$ & $1.066_{-0.063}^{+0.075}$ & \refsys{2017arXiv170306477B} \\
\end{longtable}
\begin{flushleft}
\footnotemark[1]{The new system parameters $M_{\rm s}$, $R_{\rm s}$, $\rp$, and age were derived by using the Y2 evolutionary tracks and 
the published values of stellar atmospheric parameters, stellar density from the transit fitting, and $\rp/R_{\rm s}$ (as reported in the 
corresponding reference).} \\
\footnotemark[2]{The system age was estimated by using the Y2 evolutionary tracks and 
the published values of stellar atmospheric parameters and stellar density from the transit fitting (as reported in the corresponding reference).}
\end{flushleft}
\end{landscape}
\tablebib{(\refsysbib{2008A&A...482L..17B})~\citealt{2008A&A...482L..17B} and \citealt{2012ApJ...757..161T}; 
(\refsysbib{2008A&A...482L..25B})~\citealt{2008A&A...482L..25B}; 
(\refsysbib{2008A&A...482L..21A})~\citealt{2008A&A...482L..21A}; (\refsysbib{2010A&A...511A...3G})~\citealt{2010A&A...511A...3G}; 
(\refsysbib{2009A&A...493..193L})~\citealt{2009A&A...493..193L}; (\refsysbib{2008A&A...491..889D})~\citealt{2008A&A...491..889D};
(\refsysbib{2008A&A...488L..47M})~\citealt{2008A&A...488L..47M}; (\refsysbib{2009A&A...506..281R})~\citealt{2009A&A...506..281R};
(\refsysbib{2010A&A...512A..14F})~\citealt{2010A&A...512A..14F}; (\refsysbib{2011MNRAS.417.2166S})~\citealt{2011MNRAS.417.2166S};(\refsysbib{2010A&A...520A..66B})~\citealt{2010A&A...520A..66B};
(\refsysbib{2017arXiv170306477B})~\citealt{2017arXiv170306477B}; 
(\refsysbib{2010A&A...520A..65B})~\citealt{2010A&A...520A..65B}; (\refsysbib{2010A&A...524A..55G})~\citealt{2010A&A...524A..55G};
(\refsysbib{2010A&A...520A..97G})~\citealt{2010A&A...520A..97G}; (\refsysbib{2010A&A...522A.110C})~\citealt{2010A&A...522A.110C};
(\refsysbib{2011A&A...528A..97T})~\citealt{2011A&A...528A..97T}; (\refsysbib{2012A&A...541A.149O})~\citealt{2012A&A...541A.149O};
(\refsysbib{2011A&A...531A..41C})~\citealt{2011A&A...531A..41C}; (\refsysbib{2011A&A...533A.130H})~\citealt{2011A&A...533A.130H};
(\refsysbib{2012A&A...537A.136G})~\citealt{2012A&A...537A.136G}; (\refsysbib{2012A&A...538A.145D})~\citealt{2012A&A...538A.145D}; 
(\refsysbib{2012A&A...545A...6P})~\citealt{2012A&A...545A...6P}; (\refsysbib{2012A&A...537A..54R})~\citealt{2012A&A...537A..54R};
(\refsysbib{2013A&A...555A.118A})~\citealt{2013A&A...555A.118A}; (\refsysbib{2014A&A...562A.140P})~\citealt{2014A&A...562A.140P};
(\refsysbib{2015A&A...579A..36C})~\citealt{2015A&A...579A..36C}; (\refsysbib{2008ApJ...677.1324T})~\citealt{2008ApJ...677.1324T};
(\refsysbib{2010MNRAS.401.2665P})~\citealt{2010MNRAS.401.2665P}; (\refsysbib{2008ApJ...680.1450P})~\citealt{2008ApJ...680.1450P}; 
(\refsysbib{2008ApJ...673L..79N})~\citealt{2008ApJ...673L..79N}; (\refsysbib{2013ApJ...774L..19V})~\citealt{2013ApJ...774L..19V}; 
(\refsysbib{2014A&A...570A..54L})~\citealt{2014A&A...570A..54L}; (\refsysbib{2009ApJ...704.1107L})~\citealt{2009ApJ...704.1107L}; 
(\refsysbib{2009ApJ...690.1393S})~\citealt{2009ApJ...690.1393S}; (\refsysbib{2009ApJ...706..785H})~\citealt{2009ApJ...706..785H}; 
(\refsysbib{2009ApJ...707..446B})~\citealt{2009ApJ...707..446B}; (\refsysbib{2010ApJ...715..458T})~\citealt{2010ApJ...715..458T}; 
(\refsysbib{2010ApJ...724..866K})~\citealt{2010ApJ...724..866K}; (\refsysbib{2010ApJ...720.1118B})~\citealt{2010ApJ...720.1118B};
(\refsysbib{2012ApJ...749..134H})~\citealt{2012ApJ...749..134H}; (\refsysbib{2011ApJ...726...52H})~\citealt{2011ApJ...726...52H};
(\refsysbib{2015A&A...580A..60M})~\citealt{2015A&A...580A..60M}; (\refsysbib{2011ApJ...742..116B})~\citealt{2011ApJ...742..116B};
(\refsysbib{2015A&A...577A..90M})~\citealt{2015A&A...577A..90M}; (\refsysbib{2010ApJ...725.2017K})~\citealt{2010ApJ...725.2017K}; 
(\refsysbib{2012ApJ...745...80Q})~\citealt{2012ApJ...745...80Q}; (\refsysbib{2012ApJ...760..139B})~\citealt{2012ApJ...760..139B};
(\refsysbib{2011ApJ...733..116B})~\citealt{2011ApJ...733..116B}; (\refsysbib{2011ApJ...735...24J})~\citealt{2011ApJ...735...24J};
(\refsysbib{2011AJ....142...95K})~\citealt{2011AJ....142...95K}; (\refsysbib{2011ApJ...742...59H})~\citealt{2011ApJ...742...59H};
(\refsysbib{2012AJ....144...19B})~\citealt{2012AJ....144...19B}; (\refsysbib{2012PASJ...64...97S})~\citealt{2012PASJ...64...97S};
(\refsysbib{2012AJ....144..139H})~\citealt{2012AJ....144..139H}; (\refsysbib{2013A&A...558A..86B})~\citealt{2013A&A...558A..86B};
(\refsysbib{2014AJ....147..128H})~\citealt{2014AJ....147..128H}; (\refsysbib{2014AJ....147...84B})~\citealt{2014AJ....147...84B};
(\refsysbib{2015AJ....150..168H})~\citealt{2015AJ....150..168H}; (\refsysbib{2015AJ....149..149B})~\citealt{2015AJ....149..149B};
(\refsysbib{2015PASP..127..851J})~\citealt{2015PASP..127..851J}; (\refsysbib{2015AJ....150...85H})~\citealt{2015AJ....150...85H};
(\refsysbib{2013AJ....145....5P})~\citealt{2013AJ....145....5P}; (\refsysbib{2013A&A...558A..55M})~\citealt{2013A&A...558A..55M};
(\refsysbib{2013AJ....146..113B})~\citealt{2013AJ....146..113B}; (\refsysbib{2014AJ....148...29J})~\citealt{2014AJ....148...29J};
(\refsysbib{2014AJ....147..144Z})~\citealt{2014AJ....147..144Z}; (\refsysbib{2015ApJ...813..111B})~\citealt{2015ApJ...813..111B};
(\refsysbib{2015AJ....150...49B})~\citealt{2015AJ....150...49B}; (\refsysbib{2015AJ....150...33B})~\citealt{2015AJ....150...33B};
(\refsysbib{2015A&A...580A..63M})~\citealt{2015A&A...580A..63M}; (\refsysbib{2009ApJ...696..241C})~\citealt{2009ApJ...696..241C};
(\refsysbib{2011ApJ...726....3N})~\citealt{2011ApJ...726....3N}; (\refsysbib{2008AJ....135...68H})~\citealt{2008AJ....135...68H};
(\refsysbib{2008ApJ...683L.179S})~\citealt{2008ApJ...683L.179S}; (\refsysbib{2012ApJ...756L..39B})~\citealt{2012ApJ...756L..39B};
(\refsysbib{2013ApJ...773...64P})~\citealt{2013ApJ...773...64P}; (\refsysbib{2015A&A...581L...6D})~\citealt{2015A&A...581L...6D};
(\refsysbib{2015ApJ...810...30F})~\citealt{2015ApJ...810...30F}; (\refsysbib{2010ApJ...713L.131K})~\citealt{2010ApJ...713L.131K};
(\refsysbib{2010ApJ...713L.136D})~\citealt{2010ApJ...713L.136D}; (\refsysbib{2012MNRAS.426.1291S})~\citealt{2012MNRAS.426.1291S};
(\refsysbib{2010ApJ...713L.140L})~\citealt{2010ApJ...713L.140L}; (\refsysbib{2010ApJ...724.1108J})~\citealt{2010ApJ...724.1108J};
(\refsysbib{2011ApJS..197....9F})~\citealt{2011ApJS..197....9F}; (\refsysbib{2011ApJS..197...13E})~\citealt{2011ApJS..197...13E};
(\refsysbib{2011ApJS..197...14D})~\citealt{2011ApJS..197...14D}; (\refsysbib{2012A&A...538A..96B})~\citealt{2012A&A...538A..96B};
(\refsysbib{2012A&A...547A..37B})~\citealt{2012A&A...547A..37B}; (\refsysbib{2015A&A...575A..85B})~\citealt{2015A&A...575A..85B};
(\refsysbib{2011A&A...528A..63S})~\citealt{2011A&A...528A..63S}; (\refsysbib{2012AJ....143..111J})~\citealt{2012AJ....143..111J};
(\refsysbib{2013ApJ...771...26F})~\citealt{2013ApJ...771...26F}; (\refsysbib{2013A&A...557A..74G})~\citealt{2013A&A...557A..74G};
(\refsysbib{2014A&A...564A..56D})~\citealt{2014A&A...564A..56D}; (\refsysbib{2012ApJ...761..163D})~\citealt{2012ApJ...761..163D};
(\refsysbib{2014ApJ...791...89D})~\citealt{2014ApJ...791...89D}; (\refsysbib{2014ApJ...795..151E})~\citealt{2014ApJ...795..151E};
(\refsysbib{2013ApJ...775L..11M})~\citealt{2013ApJ...775L..11M};
(\refsysbib{2015A&A...576A..11G})~\citealt{2015A&A...576A..11G}; (\refsysbib{2014A&A...572A..93H})~\citealt{2014A&A...572A..93H};
(\refsysbib{2015ApJ...803...49Q})~\citealt{2015ApJ...803...49Q}; (\refsysbib{2015A&A...575A..71A})~\citealt{2015A&A...575A..71A};
(\refsysbib{2010MNRAS.408.1689S})~\citealt{2010MNRAS.408.1689S}; (\refsysbib{2013A&A...554A..28C})~\citealt{2013A&A...554A..28C};
(\refsysbib{2015MNRAS.448.2617M})~\citealt{2015MNRAS.448.2617M}; (\refsysbib{2014MNRAS.443.2391M})~\citealt{2014MNRAS.443.2391M}; 
(\refsysbib{2009ApJ...691.1145S})~\citealt{2009ApJ...691.1145S}; (\refsysbib{2015A&A...575L..15S})~\citealt{2015A&A...575L..15S};
(\refsysbib{2011ApJ...741..114M})~\citealt{2011ApJ...741..114M}; (\refsysbib{2011ApJ...733..127S})~\citealt{2011ApJ...733..127S};
(\refsysbib{2013ApJ...779L..23P})~\citealt{2013ApJ...779L..23P}; (\refsysbib{2009MNRAS.398.1827S})~\citealt{2009MNRAS.398.1827S};
(\refsysbib{2009A&A...501..785G})~\citealt{2009A&A...501..785G}; (\refsysbib{2010A&A...517L...1Q})~\citealt{2010A&A...517L...1Q};
(\refsysbib{2015A&A...579A.136M})~\citealt{2015A&A...579A.136M}; (\refsysbib{2015MNRAS.450.1760T})~\citealt{2015MNRAS.450.1760T};
(\refsysbib{2009MNRAS.392.1532J})~\citealt{2009MNRAS.392.1532J}; (\refsysbib{2009AJ....137.4834W})~\citealt{2009AJ....137.4834W};
(\refsysbib{2009ApJ...703..752L})~\citealt{2009ApJ...703..752L}; (\refsysbib{2011MNRAS.416.2108A})~\citealt{2011MNRAS.416.2108A};
(\refsysbib{2009Natur.460.1098H})~\citealt{2009Natur.460.1098H}; (\refsysbib{2010ApJ...708..224H})~\citealt{2010ApJ...708..224H};
(\refsysbib{2015A&A...575A..61A})~\citealt{2015A&A...575A..61A}; (\refsysbib{2010A&A...519A..98B})~\citealt{2010A&A...519A..98B};
(\refsysbib{2011A&A...534A..16A})~\citealt{2011A&A...534A..16A}; (\refsysbib{2011A&A...531A..24T})~\citealt{2011A&A...531A..24T};
(\refsysbib{2010ApJ...720..337S})~\citealt{2010ApJ...720..337S}; (\refsysbib{2011MNRAS.410.1631E})~\citealt{2011MNRAS.410.1631E};
(\refsysbib{2010A&A...520A..56S})~\citealt{2010A&A...520A..56S}; (\refsysbib{2010ApJ...723L..60H})~\citealt{2010ApJ...723L..60H};
(\refsysbib{2011A&A...531A..60A})~\citealt{2011A&A...531A..60A}; (\refsysbib{2014MNRAS.440.3392B})~\citealt{2014MNRAS.440.3392B};
(\refsysbib{2011A&A...526A.130S})~\citealt{2011A&A...526A.130S}; (\refsysbib{2011AJ....142...86E})~\citealt{2011AJ....142...86E};
(\refsysbib{2012AJ....143...81S})~\citealt{2012AJ....143...81S}; (\refsysbib{2011AJ....141....8S})~\citealt{2011AJ....141....8S};
(\refsysbib{2011A&A...531A..40F})~\citealt{2011A&A...531A..40F}; (\refsysbib{2011PASP..123..547M})~\citealt{2011PASP..123..547M};
(\refsysbib{2012A&A...544A..72L})~\citealt{2012A&A...544A..72L}; (\refsysbib{2011A&A...535L...7H})~\citealt{2011A&A...535L...7H};
(\refsysbib{2012A&A...542A...4G})~\citealt{2012A&A...542A...4G};
(\refsysbib{2012MNRAS.422.1988A})~\citealt{2012MNRAS.422.1988A}; (\refsysbib{2016MNRAS.456..990C})~\citealt{2016MNRAS.456..990C};
(\refsysbib{2011A&A...533A..88G})~\citealt{2011A&A...533A..88G}; (\refsysbib{2013A&A...549A.134H})~\citealt{2013A&A...549A.134H};
(\refsysbib{2013A&A...551A..73F})~\citealt{2013A&A...551A..73F}; (\refsysbib{2012MNRAS.426..739H})~\citealt{2012MNRAS.426..739H};
(\refsysbib{2015MNRAS.454.3094S})~\citealt{2015MNRAS.454.3094S}; (\refsysbib{2013A&A...552A..82G})~\citealt{2013A&A...552A..82G};
(\refsysbib{2013A&A...559A..36G})~\citealt{2013A&A...559A..36G}; (\refsysbib{2014A&A...563A.143D})~\citealt{2014A&A...563A.143D};
(\refsysbib{2014MNRAS.445.1114A})~\citealt{2014MNRAS.445.1114A}; (\refsysbib{2013A&A...552A.120S})~\citealt{2013A&A...552A.120S};
(\refsysbib{2015AJ....150...18H})~\citealt{2015AJ....150...18H}; (\refsysbib{2013PASP..125...48M})~\citealt{2013PASP..125...48M};
(\refsysbib{2012A&A...547A..61S})~\citealt{2012A&A...547A..61S}; (\refsysbib{2013A&A...551A..80T})~\citealt{2013A&A...551A..80T};
(\refsysbib{2014A&A...572A..49N})~\citealt{2014A&A...572A..49N}; (\refsysbib{2014MNRAS.440.1982H})~\citealt{2014MNRAS.440.1982H};
(\refsysbib{2014A&A...562L...3G})~\citealt{2014A&A...562L...3G}; (\refsysbib{2014A&A...570A..64S})~\citealt{2014A&A...570A..64S};
(\refsysbib{2014A&A...568A..81L})~\citealt{2014A&A...568A..81L}; (\refsysbib{2014MNRAS.440.1470B})~\citealt{2014MNRAS.440.1470B};
(\refsysbib{2015A&A...575A.111D})~\citealt{2015A&A...575A.111D}; (\refsysbib{2015ApJ...810...11Z})~\citealt{2015ApJ...810...11Z};
(\refsysbib{2008ApJ...683.1076W})~\citealt{2008ApJ...683.1076W}; (\refsysbib{2008arXiv0805.2921M})~\citealt{2008arXiv0805.2921M};
(\refsysbib{2009ApJ...700..783P})~\citealt{2009ApJ...700..783P}; 
}


\clearpage
\newpage


\normalsize

\centering
\begin{landscape}
\begin{longtable}{l c c c c c c c c}
\caption{Derived eccentricities, RV semi-amplitudes, and trends for all the transiting giant planets considered in this study. 
Both the $1\sigma$ and $2\sigma$ upper limits are given for eccentricities consistent with zero.
The flag letters "C", "E", and "U" stand for circular, eccentric, and unconstrained orbits 
(the last concern either eccentricities compatible with zero but with uncertainties $\sigma_{e}>0.05$ or 
slightly eccentric orbits without strong enough evidence from Bayesian model comparison; see text for more details). 
"CURV" or "PLC" in place of the acceleration term indicates the presence of curvature 
regardless of its origin (distant companion, stellar activity, possible instrumental drift)
or the signal of a long-period planetary companion in the RVs.}\\
\hline 
 Planet & Flag &$e$ & $\omega$    & $e \cos{\omega} $ & $e \sin{\omega}$ & $K$                        & $\dot{\gamma}$                    & Jitter   \\
            &         &       & $[\rm deg]$  &                                &                             & $\rm [m~s^{-1}]$   & $ \rm [m~s^{-1}~day^{-1}]$  & $ \rm [m~s^{-1}]$   \\
\hline
\endfirsthead
\caption{continued.}\\
\hline 
 Planet & Flag &$e$ & $\omega$    & $e \cos{\omega} $ & $e \sin{\omega}$ & $K$                        & $\dot{\gamma}$                    & Jitter   \\
            &         &       & $[\rm deg]$  &                                &                             & $\rm [m~s^{-1}]$   & $ \rm [m~s^{-1}~day^{-1}]$  & $ \rm [m~s^{-1}]$   \\
\hline
\endhead
\hline
\endfoot
%
%
CoRoT-1b & C &
$<0.036~(<0.16)$ & 
~-~ & 
$-0.0001_{-0.0017}^{+0.0016}$ & 
$-0.0041_{-0.084}^{+0.0083}$ & 
$190\pm15$ & 
$1.03_{-0.11}^{+0.12}$ & 
$<9.1$ \\ 
CoRoT-2b & C &
$<0.024~(<0.069)$ & 
~-~ & 
$-0.00304\pm0.00055$ & 
$0.008_{-0.011}^{+0.034}$ & 
$568_{-22}^{+23}$ & 
~-~ & $40_{-10.}^{+14}$ \\ 
CoRoT-3b & C &
$<0.016~(<0.034)$ & 
~-~ & 
$-0.0057_{-0.0052}^{+0.0050}$ & 
$0.0033_{-0.0084}^{+0.016}$ & 
$2173\pm25$ & 
~-~ & $<26.0$ \\ 
CoRoT-4b & U &
$<0.14~(<0.31)$ & 
~-~ & 
$0.025_{-0.032}^{+0.038}$ & 
$-0.064_{-0.13}^{+0.069}$ & 
$62.4_{-6.7}^{+6.2}$ & 
~-~ & $<10.5$ \\ 
CoRoT-5b & U &
$<0.087~(<0.19)$ & 
~-~ & 
$-0.024_{-0.031}^{+0.024}$ & 
$-0.006_{-0.084}^{+0.058}$ & 
$61.1_{-4.4}^{+4.5}$ & 
~-~ & $<2.7$ \\ 
CoRoT-6b & U &
$<0.18~(<0.37)$ & 
~-~ & 
$-0.053_{-0.038}^{+0.039}$ & 
$0.11_{-0.10}^{+0.14}$ & 
$285_{-23}^{+26}$ & 
~-~ & $<15.8$ \\ 
CoRoT-8b & U &
$<0.19~(<0.47)$ & 
~-~ & 
$-0.044_{-0.084}^{+0.060}$ & 
$0.054_{-0.089}^{+0.21}$ & 
$27.6_{-4.7}^{+5.8}$ & 
~-~ & $<2.9$ \\ 
CoRoT-9b & E &
$0.134_{-0.034}^{+0.040}$ & 
$41_{-23}^{+30}$ & 
$0.104 \pm 0.031$ & 
$0.066 \pm 0.060$ & 
$39.0 \pm 2.4$ & 
~-~ & $<3.4$ \\
\end{longtable}
\end{landscape}




\onecolumn

\centering
\begin{longtable}{l c c c c c c}
\caption{Orbital and physical parameters of the 231 transiting giant planets considered in this work.
They were derived from our orbital fits and the system parameters listed in Table~\ref{table_system_parameters}. 
Upper limits of $1\sigma$ on the eccentricity are given for both well-determined circular orbits and undetermined eccentricities, 
as reported in the second column of Table~\ref{table_orbital_parameters} (`C' or `U').}\\
\hline 
Planet & $P$        & $a$ & $e$ &  $\mp$       & $\rho_{\rm p}$         & $\log{g_{\rm p}}$ \\
           & [days]     &  au  &        &  [$\Mjup$]  & [$\rm g\;cm^{-3}$]   & [cgs] \\
\hline
\endfirsthead
\caption{continued.}\\
\hline 
Planet & $P$        & $a$ & $e$ &  $\mp$       & $\rho_{\rm p}$         & $\log{g_{\rm p}}$ \\
           & [days]     &  au  &        &  [$\Mjup$]  & [$\rm g\;cm^{-3}$]   & [cgs] \\
\hline
\endhead
\hline
\endfoot
%
%
CoRoT-1b & $1.509$ & $0.02752_{-0.00023}^{+0.00022}$ & $<0.036$ & $1.23\pm0.10$ & $0.302_{-0.028}^{+0.031}$ & $3.015_{-0.040}^{+0.038}$ \\ 
CoRoT-2b & $1.743$ & $0.02810_{-0.00058}^{+0.00057}$ & $<0.024$ & $3.30_{-0.18}^{+0.19}$ & $1.30_{-0.13}^{+0.14}$ & $3.581\pm0.036$ \\ 
CoRoT-3b & $4.257$ & $0.0574_{-0.0013}^{+0.0012}$ & $<0.016$ & $21.44_{-0.97}^{+0.96}$ & $25.9_{-4.9}^{+6.4}$ & $4.718_{-0.063}^{+0.065}$ \\ 
CoRoT-4b & $9.202$ & $0.09025_{-0.00055}^{+0.00068}$ & $<0.14$ & $0.703_{-0.073}^{+0.071}$ & $0.517_{-0.082}^{+0.099}$ & $3.090_{-0.062}^{+0.058}$ \\ 
CoRoT-5b & $4.038$ & $0.04963_{-0.00032}^{+0.00033}$ & $<0.087$ & $0.478\pm0.035$ & $0.221_{-0.026}^{+0.030}$ & $2.788_{-0.044}^{+0.043}$ \\ 
CoRoT-6b & $8.887$ & $0.0854_{-0.0014}^{+0.0013}$ & $<0.18$ & $2.95\pm0.28$ & $2.30_{-0.29}^{+0.32}$ & $3.730_{-0.050}^{+0.048}$ \\ 
CoRoT-8b & $6.212$ & $0.06339_{-0.00098}^{+0.00096}$ & $<0.19$ & $0.218_{-0.041}^{+0.033}$ & $1.45\pm0.29$ & $3.219_{-0.093}^{+0.069}$ \\ 
CoRoT-9b & $95.273$ & $0.4029_{-0.0056}^{+0.0054}$ & $0.134_{-0.034}^{+0.040}$ & $0.846_{-0.057}^{+0.056}$ & $0.91_{-0.16}^{+0.21}$ & $3.280_{-0.060}^{+0.062}$ \\
\end{longtable}

\end{document}